\documentclass[lettersize,journal]{IEEEtran}
\usepackage{amsmath,amsfonts}
\usepackage{array}
\usepackage{textcomp}
\usepackage{stfloats}
\usepackage{url}
\usepackage{verbatim}
\usepackage{graphicx}
\usepackage{cite}
\usepackage{booktabs}
\usepackage{multirow}
\usepackage{makecell}
\usepackage{subfigure}
\usepackage[misc]{ifsym} 
\usepackage{balance}
\usepackage{listings}
\usepackage{color, xcolor}
\usepackage{algorithm2e}
\usepackage{diagbox}
\usepackage{tcolorbox}
\usepackage{stfloats}
\usepackage{enumitem}
\usepackage{cleveref}
\usepackage{ulem}
\usepackage{colortbl} 

\definecolor{lightcoral}{rgb}{0.94, 0.5, 0.5}
\definecolor{lightgreen}{rgb}{0.56, 0.93, 0.56}
\definecolor{harvestgold}{rgb}{0.85, 0.57, 0.0}
\definecolor{brightlavender}{rgb}{0.75, 0.58, 0.89}
\definecolor{capri}{rgb}{0.0, 0.75, 1.0}
\definecolor{carminepink}{rgb}{0.92, 0.3, 0.26}
\definecolor{celadon}{rgb}{0.67, 0.88, 0.69}
\definecolor{darkpastelgreen}{rgb}{0.01, 0.75, 0.24}
\definecolor{DeepSkyBlue4}{RGB}{0,104,139}

\definecolor{acccolor}{RGB}{238, 245, 252} 
\definecolor{latcolor}{RGB}{242, 242, 242} 

\begin{document}

\title{Towards Robust LLM Post-Training: Automatic Failure Management for Reinforcement Fine-Tuning}
	
\author{
	\IEEEauthorblockN{Lingzhe Zhang, Tong Jia\IEEEauthorrefmark{1}, Yunpeng Zhai, Liancheng Fang, Kening Zheng, Hongyi Liu\\ Xiaosong Huang, Philip S. Yu,~\IEEEmembership{Fellow,~IEEE} and Ying Li\IEEEauthorrefmark{1},~\IEEEmembership{Member,~IEEE}}
	\thanks{Lingzhe Zhang, Tong Jia, Hongyi Liu, Xiaosong Huang and Ying Li are with Peking University, Beijing, China. Yunpeng Zhai is with Alibaba Group, China. Liancheng Fang, Kening Zheng and Philip S. Yu are with University of Illinois Chicago, United States.}
	\thanks{Email: \{zhang.lingzhe, hxs\}@stu.pku.edu.cn, \{jia.tong, hongyiliu, li.ying\}@pku.edu.cn, zhaiyunpeng.zyp@alibaba-inc.com and \{lfang87, kzhen28, psyu\}@uic.edu}
	\thanks{* Corresponding author: Tong Jia, e-mail: (jia.tong@pku.edu.cn); Ying Li, e-mail: (li.ying@pku.edu.cn)}
}

\maketitle

\begin{abstract}
Reinforcement fine-tuning (RFT) has become a core paradigm for post-training large language models, yet its training process remains highly fragile. Existing efforts mainly improve reliability at the system level or address specific issues in individual subproblems by modifying RFT algorithms. Despite their effectiveness, they largely overlook the problem of failure management at the training-process level. When training goes wrong, practitioners still rely heavily on expert-driven manual inspection and correction, and automatic failure management for RFT remains largely unexplored. In this paper, we take a first step toward systematic failure management for reinforcement fine-tuning. To understand the empirical structure of RFT failures, we first construct RFT-FaultBench, the first benchmark for fine-grained failures in reinforcement fine-tuning, covering 5 fault families, 16 fault types, 779 training runs, 22,549 train-step records, and 1,457,288 trajectory-level records. Based on this benchmark, we conduct a comprehensive empirical study showing that RFT failures are both observable from training dynamics and distinguishable through their empirical fault fingerprints. Building on these findings, we propose RFT-FM, an automatic failure management framework for reinforcement fine-tuning that unifies anomaly detection, failure diagnosis, and atuo remediation in a closed loop. Experimental results show that RFT-FaultBench is neither trivial nor saturated: it exhibits clear anomaly structure while still posing substantial challenges, especially under subtle fault settings. Moreover, RFT-FM shows strong capability in detecting, diagnosing, and mitigating RFT failures.
\end{abstract}


\section{Introduction}

Large language models (LLMs) have achieved remarkable progress in recent years, demonstrating strong capabilities in language understanding, reasoning, coding, and decision-making across diverse domains~\cite{guo2025deepseek, el2025competitive, zhang2025survey2, pan2025omni, pan2025d}. Beyond large-scale pretraining, much of this progress now depends on \emph{post-training}, which aligns pretrained models with task objectives, human preferences, and downstream interaction requirements. In particular, reinforcement fine-tuning (RFT), including RLHF-style post-training, has become an increasingly important stage for improving reasoning quality, response behavior, and goal-directed performance.

However, RFT is also one of the most fragile stages in the LLM development pipeline. Unlike standard supervised fine-tuning, reinforcement-based post-training couples policy generation, reward assignment, constrained optimization, credit assignment, and sometimes tool or environment interaction into a tightly interdependent learning process. Failures in any of these components can distort the training dynamics, degrade model behavior, or even silently invalidate the optimization objective.

To address these challenges, a growing body of work has been proposed from multiple perspectives. Some studies focus on system- and infrastructure-level reliability in large-scale LLM training. For example, L4~\cite{jiang2025l4} diagnoses large-scale LLM training failures by exploiting cross-job, spatial, and temporal log patterns to localize faulty nodes and failure-indicating logs. RobustRL~\cite{chen2025role} improves the Effective Training Time Ratio (ETTR) of RL post-training under GPU machine failures through a Detect-Restart-Reconnect paradigm. Other system efforts such as TRANSOM~\cite{wu2023transom} and FlashRecovery~\cite{zhang2025flashrecovery} further improve failure recovery efficiency through automated monitoring, restart orchestration, and low-cost restoration mechanisms. Another line of work focuses on algorithmic stabilization and reward-aware mitigation within RLHF itself. For instance, prior studies have investigated reward shaping and reward-model redesign to reduce reward hacking~\cite{fu2025reward, miao2025information, duan2026mitigating}, while other work revisits KL regularization and stable optimization in RLHF to better balance alignment quality and training stability~\cite{liu2025rethinking, heunifying}.

\begin{figure}[htbp]
	\centering
	\includegraphics[width=1\linewidth]{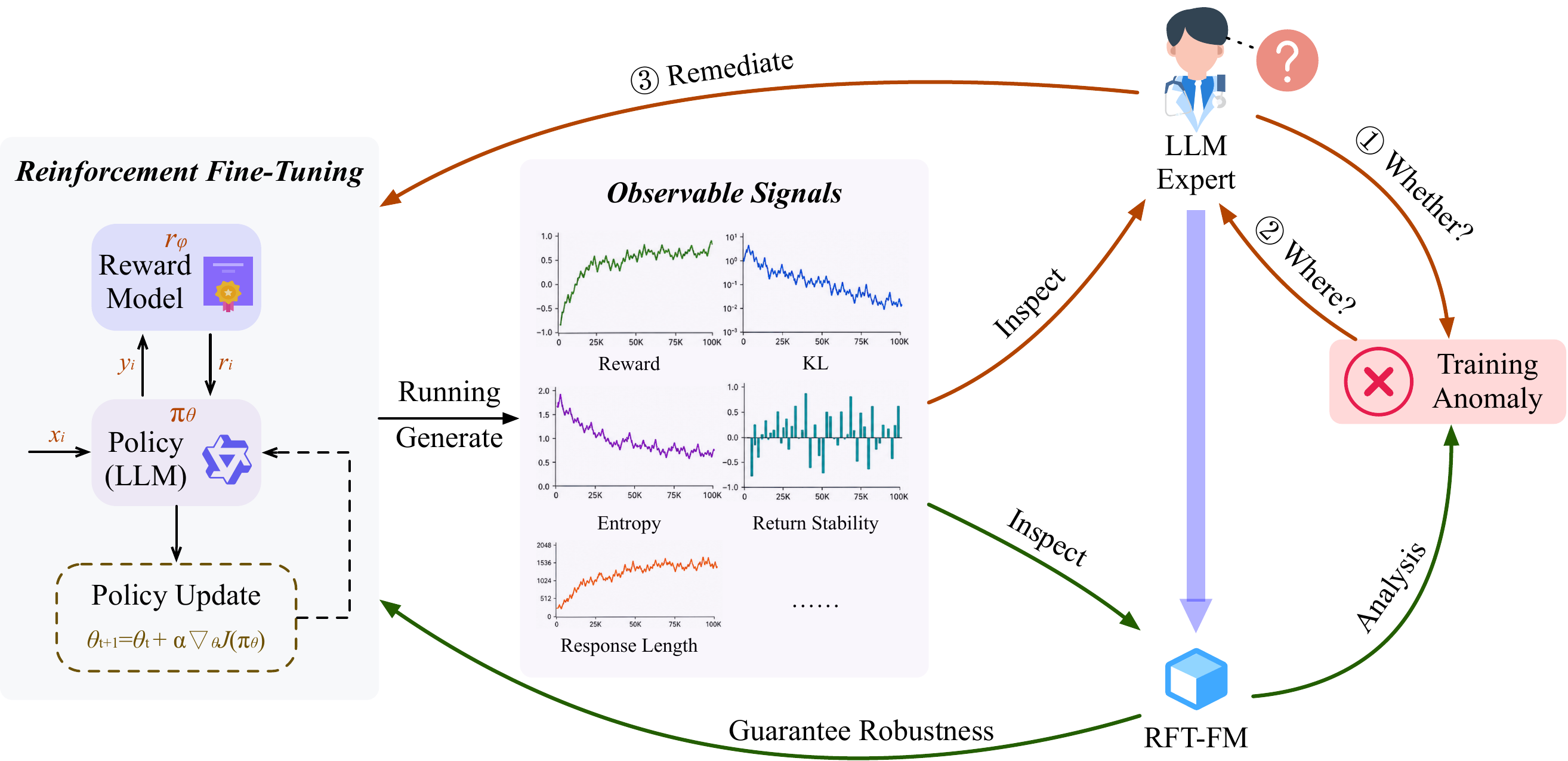}
	\caption{Training Anomalies in Reinforcement Fine-Tuning: From Manual Inspection to Automatic Failure Management.}
	\label{fig: intro-example}
\end{figure}

While these efforts substantially improve the robustness of reinforcement fine-tuning from their respective perspectives, \uline{failure handling in RFT remains largely ad hoc in practice}. When training goes wrong, LLM experts still need to manually inspect training curves, logs, and sampled outputs, infer possible root causes, adjust hyperparameters or system settings, and rerun the experiment. Such workflows are labor-intensive, experience-dependent, and difficult to scale, especially as post-training pipelines become increasingly complex and expensive. As illustrated in Fig.~\ref{fig: intro-example}, this gap motivates a broader vision of \uline{Automatic Failure Management for RFT}, in which failure analysis, diagnosis, and downstream remediation can be carried out with substantially reduced reliance on human experts. To realize this vision, two key challenges remain.

\begin{itemize}[leftmargin=*]
	\item \textbf{Data-level challenge.} It remains unclear whether failures in RFT exhibit sufficiently stable, recurring, and distinguishable patterns to be studied as training anomalies in the first place. Without establishing such empirical structure, it is difficult to determine what constitutes an anomaly, how different failures should be categorized, or what data foundation is needed for systematic study. Moreover, there is currently no established benchmark dataset for fine-grained RFT training anomalies to support such investigation.
	
	\item \textbf{Method-level challenge.} Even if such anomalies can be empirically established, designing effective methods for them remains nontrivial. The intermediate signals produced during RFT differ fundamentally from the run-time data typically used in existing failure management settings. Rather than relying primarily on infrastructure metrics or application logs, RFT exposes a tightly coupled observability surface spanning reward trajectories, KL dynamics, entropy evolution, return stability, generation behavior, and tool/environment feedback. As a result, existing software-oriented failure management methods~\cite{zhang2025survey, notaro2021survey, remil2024aiops} cannot be directly transferred without rethinking how anomaly signals should be represented, aggregated, and diagnosed in post-training dynamics.
\end{itemize}

\uline{To address the data-level challenge, we a benchmark dataset for fine-grained training anomalies in reinforcement fine-tuning (named RFT-FaultBench), and use it to systematically study their patterns, structure, and distinguishability.} \textbf{RFT-FaultBench} provides broad and structured coverage of RFT failures. It includes 16 fine-grained fault types across 5 fault families, with both standard and hardmode variants to capture anomalies ranging from obvious failures to weaker and more realistic perturbations. Overall, the benchmark contains 779 training runs, including 320 standard fault runs, 448 hardmode fault runs, and 11 normal baseline runs, together with more than 26,000 train-step records and rich trajectory-level artifacts.

Building on RFT-FaultBench, we conduct an empirical study to examine whether RFT failures form a meaningful anomaly space. Specifically, we analyze their patterns, consistency, and separability across fault families, severity levels, and observable training signals. Our results indicate that RFT failures are not arbitrary noise; instead, they induce stable and structured anomaly signatures that can support systematic detection and diagnosis.

\uline{To address the method-level challenge, we propose RFT-FM, an automatic failure management framework for reinforcement fine-tuning.} \textbf{RFT-FM} consists of three tightly coupled components: \textit{RFT-Feature-Based IVS Scoring} for anomaly detection, \textit{Training-Dynamics Failure Attribution} for failure diagnosis, and \textit{Agentic Training Intervention} for auto remediation. Operating directly on post-training dynamics such as reward, KL, entropy, returns, generation behavior, and tool/environment feedback, RFT-FM turns failure handling in RFT from manual expert inspection into a structured and automatable process.

We evaluate the effectiveness of RFT-FM on RFT-FaultBench. The results show that, in terms of anomaly detection, RFT-FM achieves an F1 score of 87.96\% under the easy setting and 73.88\% under the hard setting. In terms of failure attribution, it reaches a type-level Macro-F1 of 85.51\% on easy faults and 42.16\% on hard faults, showing that different RFT failures are distinguishable through their training-dynamics fingerprints while still remaining challenging under subtle anomaly settings. In terms of automatic remediation, RFT-FM further provides a preliminary closed-loop intervention capability, achieving a mitigation rate of 46.25\% across remediation cases. In summary, the key contributions of this work are as follows:

\begin{itemize}[leftmargin=*]
	\item We construct and release the first benchmark dataset for fine-grained failures in reinforcement fine-tuning, namely RFT-FaultBench, covering 5 fault families, 16 fault types, 779 training runs, 22,549 train-step records, and 1,457,288 trajectory-level records across both easy and hard anomaly settings.
	\item Based on this dataset, we conduct a comprehensive empirical study on RFT training anomalies, showing that RFT failures are not only observable from training dynamics but also distinguishable through their empirical fault fingerprints.
	\item Building on these findings, we propose RFT-FM, an automatic failure management framework for reinforcement fine-tuning, which unifies RFT-specific anomaly detection, failure diagnosis, and automatic remediation within a closed-loop pipeline.
	\item Extensive experiments demonstrate that RFT-FaultBench is neither trivial nor saturated: it exhibits clear anomaly structure while still posing substantial challenges, especially under subtle fault settings. Moreover, RFT-FM shows strong capability in detecting, diagnosing, and mitigating RFT failures.
\end{itemize}

\section{Background}

\subsection{Reinforcement Fine-Tuning}

Reinforcement Fine-Tuning (RFT) is a paradigm that leverages reinforcement learning (RL) to adapt large language models to complex decision-making tasks in a reward-driven manner~\cite{christiano2017deep, ziegler2019fine}. This approach enables the fine-tuning of models using reinforcement learning techniques to enhance their performance on specialized tasks with minimal training data.

Unlike supervised fine-tuning (SFT), which requires high-quality labeled data and learns via direct instruction, RFT allows models to explore action spaces and learn from reward signals, enabling adaptation in settings where ground truth annotations are scarce or ambiguous. This makes RFT particularly suitable for tasks such as code generation, multi-step reasoning, and tool-use planning.

A number of core algorithms underpin modern RFT strategies. Direct Preference Optimization (DPO)~\cite{rafailov2023direct} simplifies the traditional reward model plus Proximal Policy Optimization (PPO) pipeline by directly optimizing preference-based objectives. PPO~\cite{schulman2017proximal} remains widely used due to its stability in continuous action spaces. Group Relative Policy Optimization (GRPO)~\cite{shao2024deepseekmath} extends PPO by introducing group-based policy updates, allowing for more robust modeling of action diversity.

From a training-systems perspective, modern RFT typically involves several tightly coupled components, including rollout generation, reward computation, reference-constrained policy updates, value estimation, and sometimes tool or environment interaction. During this process, the training system naturally exposes a variety of intermediate signals, such as reward trajectories, KL divergence, entropy, returns, response length, and policy loss. These signals provide an important view into the internal dynamics of post-training.

At the same time, this coupling also makes RFT substantially more fragile than standard supervised fine-tuning. Errors in reward design, unstable policy updates, misaligned credit assignment, or corrupted interaction feedback can all propagate through training and distort the optimization process. As a result, RFT failures often manifest not only in final model outputs, but also in the training dynamics themselves.

\begin{table*}[t]
	\centering
	\caption{Taxonomy of RFT Training Anomalies}
	\label{tab: fault_taxonomy}
	\begin{tabular}{lllp{9.2cm}}
		\toprule
		\textbf{Family} & \textbf{Fault ID} & \textbf{Fault Name} & \textbf{Description} \\
		\midrule
		\multirow{3}{*}{Reward Fault}
		& RF-1  & Reward Spike
		& Reward values are inflated beyond normal quality-aligned ranges. \\
		& RF-2  & Reward Collapse
		& Reward signals are suppressed or driven toward near-zero values. \\
		& RF-3  & Reward Hacking
		& The policy exploits reward flaws without truly task solving. \\
		
		\midrule
		\multirow{4}{*}{Policy Generation Fault}
		& PG-1  & Empty Response
		& The policy produces blank, trivial, or nearly content-free outputs. \\
		& PG-2  & Repetition Collapse
		& The policy degenerates into repeated tokens, phrases, or loops. \\
		& PG-3S & Length Short
		& Responses are abnormally short relative to task expectations. \\
		& PG-3L & Length Long
		& Responses are abnormally long, trailing, or weakly controlled. \\
		
		\midrule
		\multirow{3}{*}{Optimization Dynamics Fault}
		& OD-1  & KL Explosion
		& Policy updates become aggressive and diverge from the reference. \\
		& OD-2  & Update Freeze
		& Effective policy updates nearly vanish while training still runs. \\
		& OD-3  & Entropy Collapse
		& Policy entropy drops too quickly, causing exploration collapse. \\
		
		\midrule
		\multirow{3}{*}{Credit Assignment Fault}
		& CA-1  & Value Mismatch
		& Value estimates remain inconsistent with realized returns. \\
		& CA-2  & Advantage Instability
		& Advantage estimates become excessively noisy or unstable. \\
		& CA-3  & Delayed Credit
		& Delayed reward feedback is misattributed across actions or steps. \\
		
		\midrule
		\multirow{3}{*}{Tool/Environment Fault}
		& TE-1  & Tool Call Error
		& Tool interactions fail due to malformed calls or execution errors. \\
		& TE-2  & Observation Corruption
		& Environment observations are corrupted before reaching policy. \\
		& TE-3  & Termination Error
		& Episodes end incorrectly or prematurely, truncating rollouts. \\
		
		\bottomrule
	\end{tabular}
\end{table*}

\subsection{Software Failure Management}

Software failure management is a general workflow for maintaining system reliability in the presence of runtime failures, degradations, and abnormal behaviors. It is commonly organized into three stages: anomaly detection, failure diagnosis, and automated remediation.

Anomaly detection serves as the first stage of the workflow and is responsible for identifying deviations from normal system behavior during execution. Once an anomaly is detected within a specific time window or operational context, the process proceeds to failure diagnosis. Failure diagnosis aims to localize the source of the detected failure and infer its underlying cause. In practice, this stage is often formulated as a classification, attribution, or root-cause analysis task, depending on the granularity of the failure space and the available observability signals.

Based on the diagnosis result, the system then performs automated remediation. This stage applies corrective or mitigating actions to restore normal behavior, reduce the impact of the failure, or prevent recurrence. Typical remediation actions may include configuration adjustment, component restart, workload migration, rollback, or other intervention mechanisms, depending on the target system.

Although failure management in reinforcement fine-tuning differs substantially from conventional software settings in its observability signals, fault space, and intervention space, the overall workflow remains analogous. In both cases, the system must first detect abnormal behavior, then determine what has gone wrong, and finally trigger targeted recovery actions. This shared process perspective motivates treating RFT failure handling as a structured failure-management problem rather than as purely ad hoc debugging.

\section{Dataset Construction}

To investigate whether failures in reinforcement fine-tuning exhibit sufficiently stable, recurring, and distinguishable patterns, we first construct a dedicated benchmark dataset of RFT training anomalies. The benchmark is built on top of the OpenRLHF~\cite{hu2024openrlhf} training framework and instantiated on a controlled arithmetic-style reasoning task, providing a reproducible environment for systematic fault injection and anomaly analysis. All runs are collected under matched fault/control settings in an 8-GPU training configuration. The experiments are conducted on a server running Ubuntu 22.04.5 LTS, with 8 NVIDIA H20 GPUs, 160 Intel Xeon CPU cores across two sockets, 1.8 TiB RAM, and multi-terabyte NVMe storage.

\subsection{RFT Training Anomalies}

To study failure management in reinforcement fine-tuning, we first construct a taxonomy of representative RFT training anomalies. In doing so, we draw on prior work showing that RFT and RLHF can fail in recurring ways, including reward hacking, preference collapse under optimization bias, and behavior-level degeneration after reinforcement-based post-training~\cite{fu2025reward, wenlanguage, chen2024odin, miao2025information, ji2025language, sheshadrisome, xue2026supervised, ouyang2022training}. Based on these observations, we identify 16 fine-grained fault types spanning 5 fault families. Together, these fault types cover a broad range of anomaly patterns that are both practically meaningful and amenable to controlled injection. While real-world RFT systems may exhibit more diverse and entangled failures than those enumerated here, the selected fault types capture many representative and diagnostically relevant cases for benchmark construction.

As shown in Table~\ref{tab: fault_taxonomy}, the resulting taxonomy covers five major fault families in RFT. For Reward Faults (RF), we include Reward Spike (RF-1), Reward Collapse (RF-2), and Reward Hacking (RF-3), which capture abnormal reward inflation, reward suppression, and reward--objective mismatch, respectively. For Policy Generation Faults (PG), we include Empty Response (PG-1), Repetition Collapse (PG-2), Length Short (PG-3S), and Length Long (PG-3L), covering common degeneration patterns in rollout generation. For Optimization Dynamics Faults (OD), we include KL Explosion (OD-1), Update Freeze (OD-2), and Entropy Collapse (OD-3), which reflect unstable or degenerate policy updates during post-training. For Credit Assignment Faults (CA), we include Value Mismatch (CA-1), Advantage Instability (CA-2), and Delayed Credit (CA-3), targeting failures in value estimation and reward propagation. Finally, for Tool/Environment Faults (TE), we include Tool Call Error (TE-1), Observation Corruption (TE-2), and Termination Error (TE-3), which represent interaction failures between the policy and its external execution context.

\begin{table*}[htbp]
	\centering
	\caption{Family-Level Statistics of RFT-FaultBench}
	\label{tab: dataset_family_summary}
	\setlength{\tabcolsep}{11.7pt}
	\begin{tabular}{llrrrrrrrrr}
		\toprule
		\multirow{2}{*}{\textbf{Family}} & \multirow{2}{*}{\textbf{\# Types}}
		& \multicolumn{3}{c}{\textbf{Runs}}
		& \multicolumn{3}{c}{\textbf{Train-Step Records}}
		& \multicolumn{3}{c}{\textbf{Trajectory Records}} \\
		\cmidrule(lr){3-5} \cmidrule(lr){6-8} \cmidrule(lr){9-11}
		&  & \textbf{Easy} & \textbf{Hard} & \textbf{Total}
		& \textbf{Easy} & \textbf{Hard} & \textbf{Total}
		& \textbf{Easy} & \textbf{Hard} & \textbf{Total} \\
		\midrule
		RF     & 3  & 60 & 84  & 144 & 1,200 & 3,360 & 4,560 & 76,800  & 215,040 & 291,840 \\
		PG     & 4  & 80 & 112 & 192 & 1,585 & 3,880 & 5,465 & 101,448 & 248,320 & 349,768 \\
		OD     & 3  & 60 & 84  & 144 & 1,204 & 3,220 & 4,424 & 77,120  & 206,080 & 283,200 \\
		CA     & 3  & 60 & 84  & 144 & 1,200 & 3,360 & 4,560 & 76,800  & 215,040 & 291,840 \\
		TE     & 3  & 60 & 84  & 144 & 1,200 & 2,340 & 3,540 & 76,800  & 149,760 & 226,560 \\
		Normal & -- & -- & --  & 11  & --    & --    & 220   & --      & --      & 14,080 \\
		\midrule
		\textbf{Total} & \textbf{16}
		& \textbf{320} & \textbf{448} & \textbf{779}
		& \textbf{6,389} & \textbf{16,160} & \textbf{22,549}
		& \textbf{408,968} & \textbf{1,034,240} & \textbf{1,457,288} \\
		\bottomrule
	\end{tabular}
\end{table*}

\subsection{Anomaly Injection}

\begin{figure}[htbp]
	\centering
	\includegraphics[width=1\linewidth]{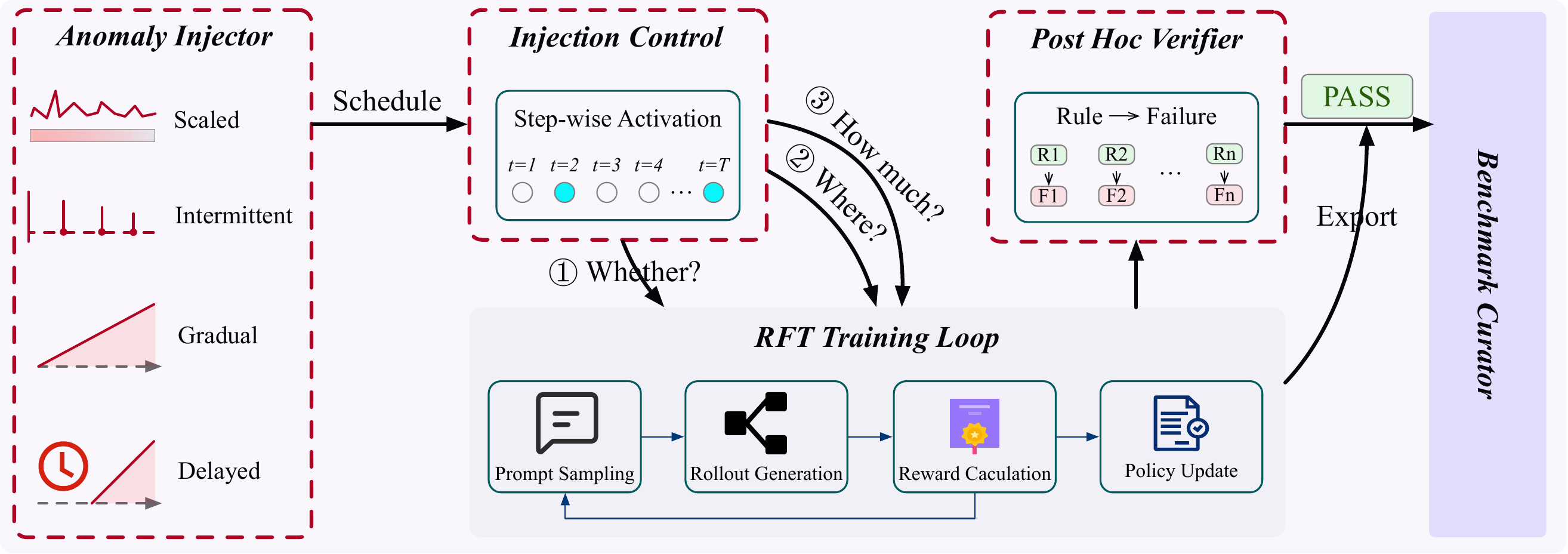}
	\caption{Architecture of RFT-Specific Anomaly Injection}
	\label{fig: anomaly-injection-architecture}
\end{figure}

To simulate RFT training failures in a more diverse, multi-scale, and realistic manner, and to ensure that injected faults indeed produce their intended anomaly signatures, we design an RFT-specific anomaly injection framework. As shown in Fig.~\ref{fig: anomaly-injection-architecture}, the framework is centered around an \emph{Injection Control} module, which schedules and parameterizes the \emph{Anomaly Injector} during training. Rather than applying only fixed perturbations, the controller regulates anomaly activation through multiple dimensions, including scaled strength, intermittent triggering, gradual ramp-up, and delayed onset. This design allows the same fault type to appear with different magnitudes, temporal patterns, and observability levels, thereby producing more realistic and challenging anomaly behaviors in the training process.

The Anomaly Injector interfaces directly with the RFT training loop and perturbs concrete stages of post-training, including rollout generation, reward computation, policy/value updates, and tool or environment interaction. In this way, anomaly injection is performed online inside the training process, rather than as a post hoc corruption over saved outputs. As training proceeds, the perturbed runs naturally produce observable deviations in telemetry such as reward, KL, entropy, returns, response length, and policy loss, which are then exported for downstream analysis.

To ensure benchmark quality, the injected runs are passed to a Post Hoc Verifier after training. This verifier applies fault-specific checking rules to determine whether the intended anomaly has been successfully induced and whether the resulting telemetry matches the expected fault signature. Only runs that pass this verification stage are retained by the Benchmark Curator, which organizes the verified artifacts, metadata, and train-step records into the final benchmark dataset. Through this injection--verification--curation pipeline, the benchmark construction process maintains both controllability and realism, while ensuring that the retained anomalies are structurally meaningful for subsequent failure management tasks.

\subsection{Dataset Details}

Based on the above anomaly taxonomy and injection pipeline, we construct \textsc{RFT-FaultBench}, a benchmark dataset for fine-grained training anomalies in reinforcement fine-tuning. The dataset is organized along multiple dimensions. Semantically, it covers 5 fault families and 16 fine-grained fault types. Operationally, it contains 779 training runs in total, including 768 fault-injected runs and 11 normal baseline runs. Structurally, each run is paired with exported train-step telemetry, trajectory-level records, fault injection metadata, and post hoc verification results. In aggregate, \textsc{RFT-FaultBench} contains 22,549 train-step records and 1,457,288 trajectory-level records. 

In addition, the dataset is organized into two difficulty regimes, namely easy and hard. Easy anomalies correspond to more salient failure manifestations, whereas hard anomalies preserve the same underlying fault semantics but exhibit weaker, more localized, or more subtle signatures in training dynamics. This design enables evaluation not only under clearly visible failures, but also under more challenging anomaly settings with reduced observability and separability. Table~\ref{tab: dataset_family_summary} summarizes the family-level statistics across difficulty regimes. We release the dataset, code, and documentation at: \url{https://github.com/AIOps4LLM/RFT-FaultBench}.

\section{Empirical Study}

Based on RFT-FaultBench, we conduct an empirical study to investigate whether failures in reinforcement fine-tuning form a meaningful and structured anomaly space in training dynamics. Our study is organized around the following two research questions:

\begin{itemize}[leftmargin=*]
	\item \textbf{RQ1:} Can failures in reinforcement fine-tuning be consistently observed as anomalies from training dynamics?
	\item \textbf{RQ2:} Do different RFT failures induce stable and distinguishable divergence patterns in training dynamics?
\end{itemize}

\subsection{Observability}

We first study whether failures in reinforcement fine-tuning are observable as anomalies from training dynamics. The key question is whether faulty training runs exhibit stable deviations from healthy optimization trajectories, rather than merely producing isolated or noisy irregularities. To this end, we compare normal and faulty runs across core RFT telemetry, including reward, KL divergence, entropy, and response length, and examine whether different fault types produce persistent and repeatable anomaly signals. For visual clarity, we focus here on representative faults from the easy setting, selecting one visually salient fault from each family to illustrate their characteristic anomaly patterns.

\begin{figure}[htbp]
	\centering
	\includegraphics[width=1\linewidth]{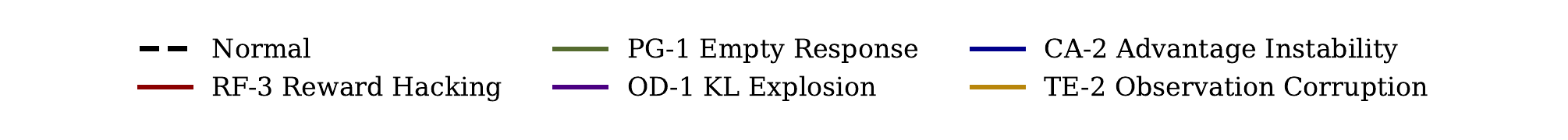}
	\vspace{0.4em}
	\subfigure[Reward]{
		\begin{minipage}{0.46\linewidth}
			\centering
			\includegraphics[width=\linewidth]{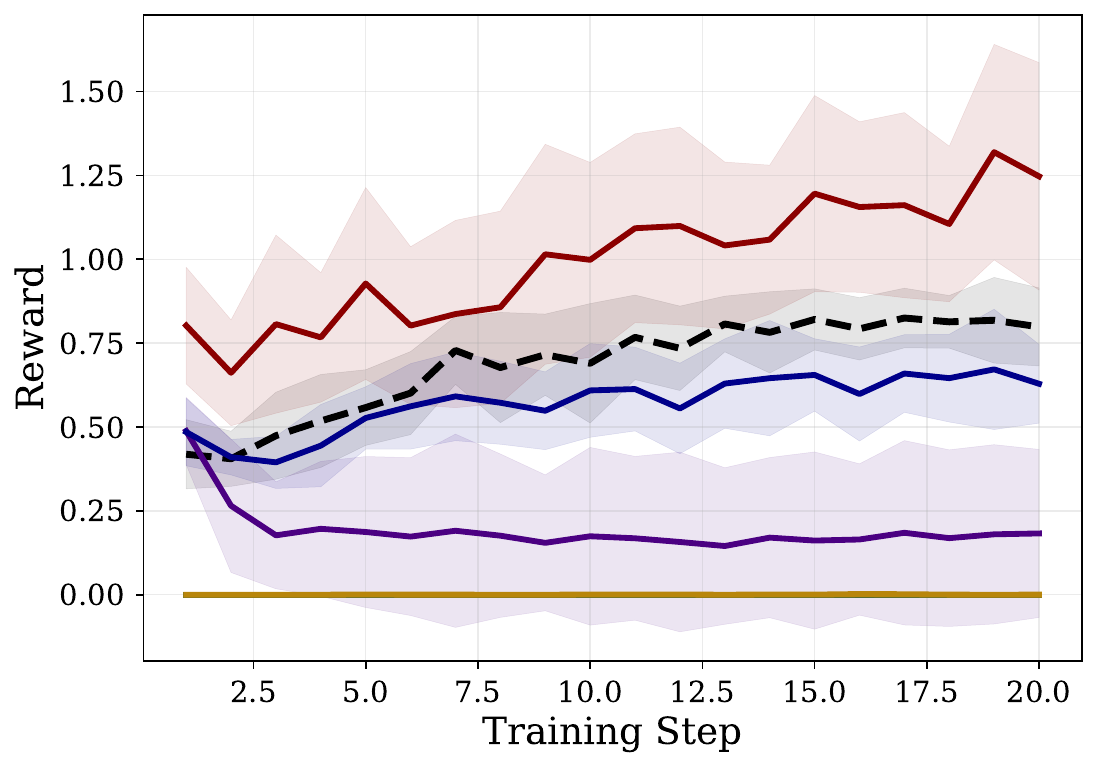}
			\label{fig:observability-reward}
		\end{minipage}
	}
	\subfigure[KL Divergence]{
		\begin{minipage}{0.46\linewidth}
			\centering
			\includegraphics[width=\linewidth]{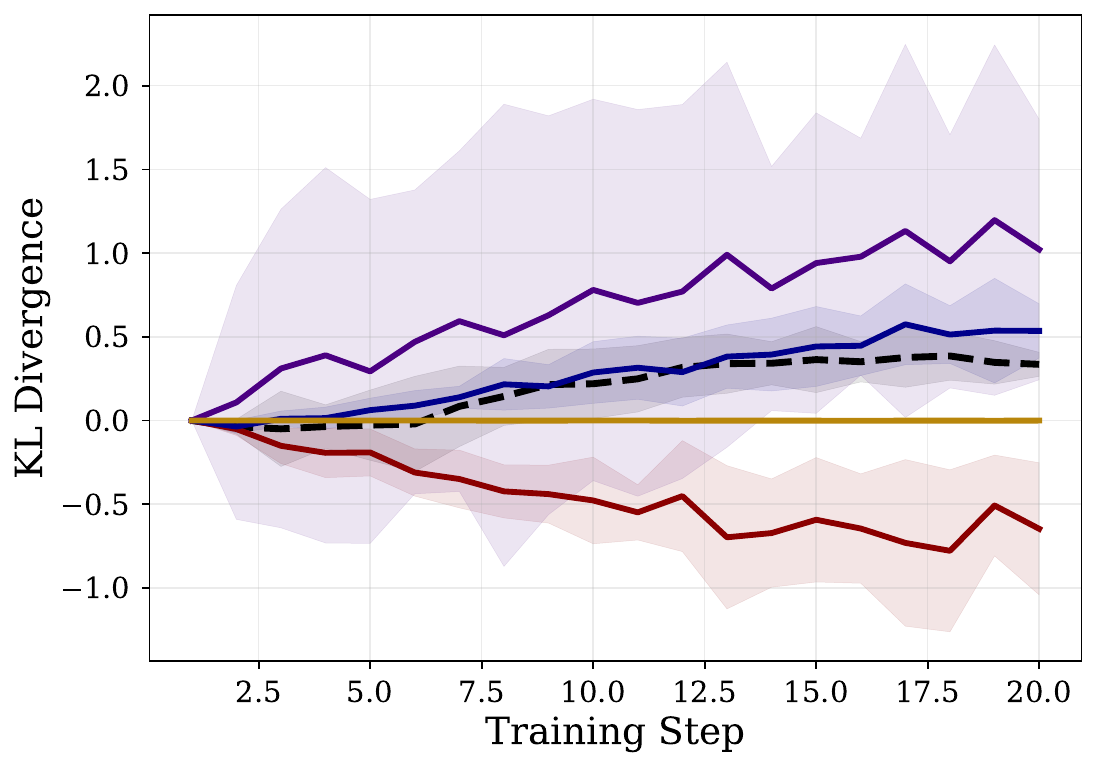}
			\label{fig:observability-kl}
		\end{minipage}
	}
	\subfigure[Entropy]{
		\begin{minipage}{0.46\linewidth}
			\centering
			\includegraphics[width=\linewidth]{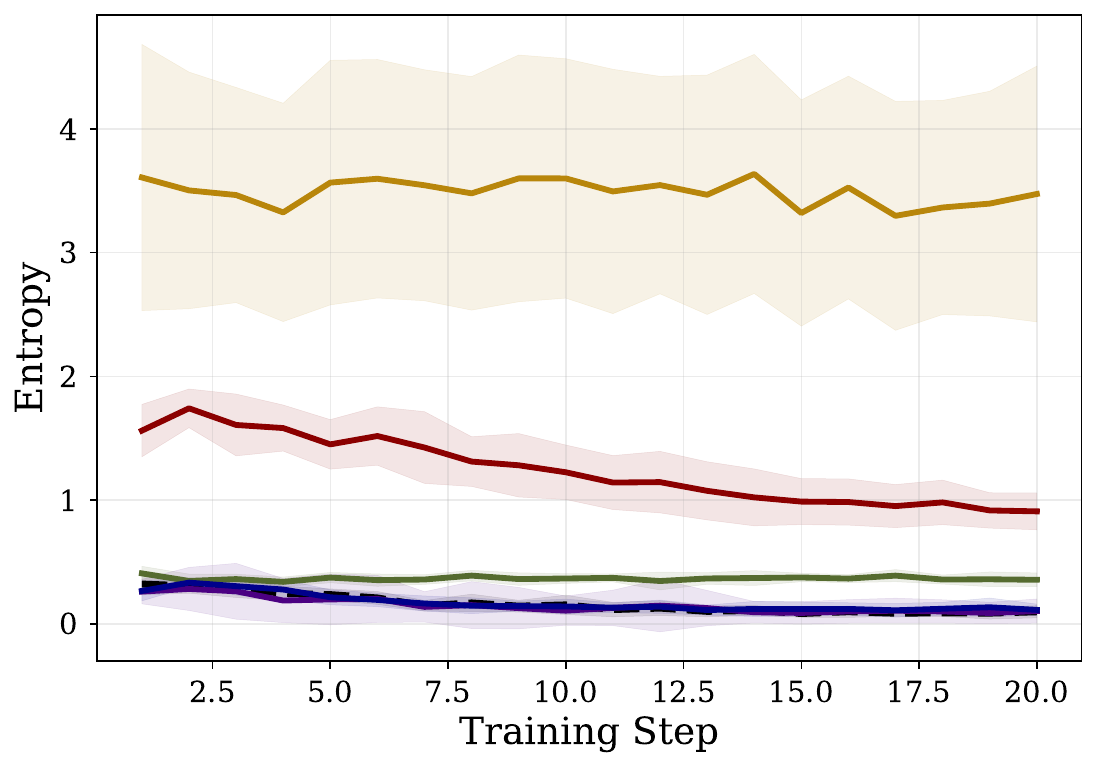}
			\label{fig:observability-entropy}
		\end{minipage}
	}
	\subfigure[Response Length]{
		\begin{minipage}{0.46\linewidth}
			\centering
			\includegraphics[width=\linewidth]{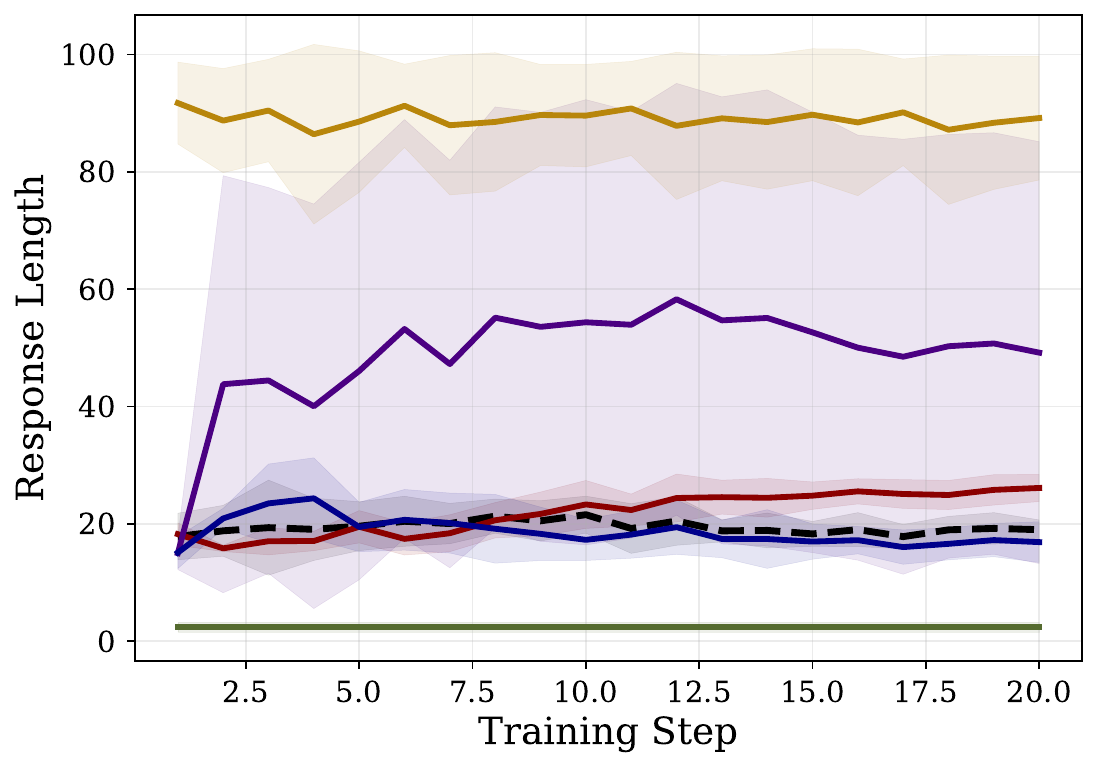}
			\label{fig:observability-length}
		\end{minipage}
	}
	
	\caption{Observability of RFT Failures in Training Dynamics}
	\label{fig:observability}
\end{figure}

Fig.~\ref{fig:observability} shows representative training trajectories of healthy runs and five fault families. A normal run exhibits a relatively coherent optimization pattern: reward improves steadily, KL divergence remains bounded, entropy gradually decreases, and response length evolves within a comparatively stable range. In contrast, faulty runs produce persistent deviations from this healthy trajectory, indicating that RFT failures leave observable footprints throughout training.

At the same time, the figure also shows that observability is highly heterogeneous across fault families. Reward faults such as RF-3 shift the reward trajectory upward while simultaneously introducing abnormal optimization behavior, reflecting a mismatch between apparent reward improvement and actual training quality. Policy-generation faults such as PG-1 lead to near-collapsed reward and response length, together with frozen or weakly evolving optimization signals. Optimization faults such as OD-1 are particularly salient in KL divergence and response length, where the deviation from normal training becomes increasingly pronounced over time. Tool/environment faults such as TE-2 exhibit large disruptions across multiple signals simultaneously, suggesting that external interaction failures can propagate broadly into the training dynamics.

Compared with these strongly visible failures, credit-assignment faults are much closer to healthy runs in coarse-grained telemetry. In Fig.~\ref{fig:observability}, the representative CA-2 case does not exhibit the same level of first-order separation as PG, OD, or TE faults in reward, KL, entropy, or response length alone. This suggests that observability in RFT is not uniform: some failures manifest as prominent single-signal deviations, whereas others appear only as weaker and more structured inconsistencies across training signals. In other words, RFT anomalies are observable, but not all of them are equally salient under coarse human-visible views.

Overall, these results support an affirmative answer to RQ1. Failures in reinforcement fine-tuning can indeed be observed as anomalies from training dynamics, and their signatures are typically persistent rather than incidental. However, the strength of observability varies substantially across fault types, motivating more systematic and structured anomaly analysis beyond simple inspection of individual metrics.

\begin{center}
	\setlength{\fboxsep}{5pt}
	\noindent\fcolorbox{black}{gray!10}{
		\begin{minipage}{0.93\linewidth}
			\textbf{Summary.} RFT failures are empirically observable from training dynamics, but their observability is inherently heterogeneous. Some failures produce clear first-order deviations, while others remain subtle and require more structured analysis.
		\end{minipage}
	}
\end{center}

\subsection{Distinguishability}

Observability alone does not imply that different RFT failures are meaningfully distinguishable from one another. A structured anomaly space further requires that different failure sources induce relatively stable yet different deviation patterns, rather than collapsing into a single notion of abnormal training. We therefore next study whether different RFT failures exhibit distinguishable signatures in training dynamics.

To answer this question, we summarize each fault type using a set of empirical training-dynamics signatures computed from observed reward, KL divergence, entropy, and response-length trajectories, and compare how different faults occupy the resulting signature space.

\begin{figure}[htbp]
	\centering
	\includegraphics[width=1\linewidth]{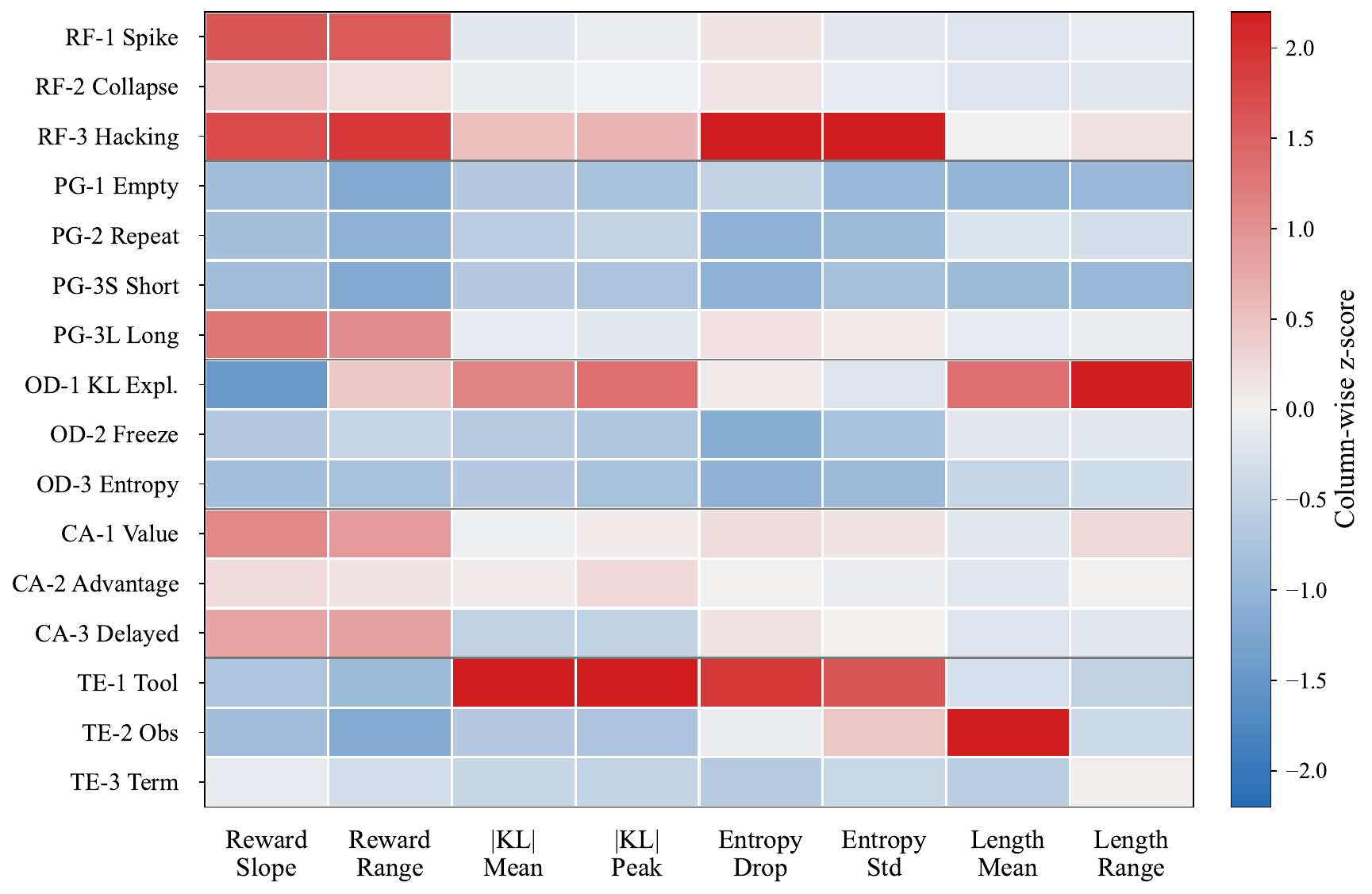}
	\caption{Empirical Fingerprints of Different RFT Failures}
	\label{fig: distinguishability-heatmap}
\end{figure}

Fig.~\ref{fig: distinguishability-heatmap} shows that different RFT failures indeed exhibit different empirical fingerprints, rather than differing only in anomaly magnitude. Reward faults are primarily characterized by reward-side deviations: RF-1 and RF-2 mainly alter reward slope and reward range, whereas RF-3 further introduces substantially stronger entropy-side deviations, making it structurally different from the other reward faults. Policy-generation faults occupy another region of the signature space. PG-1, PG-2, and PG-3S are all associated with suppressed reward-side behavior, but they differ in the extent to which generation length and entropy-related signatures are affected. In contrast, PG-3L is clearly separated from these collapse-style generation faults, with much stronger length-related signatures and comparatively weaker collapse behavior.

Optimization faults are dominated by KL- and length-related deviations. In particular, OD-1 exhibits strong peaks in both KL and response-length signatures, making it one of the most distinctive fault types in the entire benchmark. OD-2 and OD-3 remain closer to one another, but still differ from policy-generation and reward faults in the relative emphasis placed on optimization-side signals. Tool/environment faults form another distinguishable pattern family. TE-1 is characterized by strong KL- and entropy-related disruptions, whereas TE-2 is especially prominent along the response-length dimension, indicating that external interaction failures may propagate into training dynamics in qualitatively different ways.

Compared with these more salient cases, credit-assignment faults are noticeably subtler. CA-1, CA-2, and CA-3 remain closer to the center of the signature space and exhibit weaker first-order deviations across several coarse summary dimensions. However, they do not collapse completely into the normal pattern or into any single non-CA family. Instead, they occupy a relatively compact yet still distinct region, suggesting that weakly observable failures may still be distinguishable when summarized through multi-signal training-dynamics signatures rather than through any single metric alone.

Overall, these results support an affirmative answer to RQ2. Different RFT failures do not merely correspond to a generic state of abnormal training; instead, they induce heterogeneous and structured anomaly signatures in training dynamics. Although the degree of separability is uneven across fault types, the empirical signature space is sufficiently structured to show that RFT failures are not only observable, but also distinguishable.

\begin{center}
	\setlength{\fboxsep}{5pt}
	\noindent\fcolorbox{black}{gray!10}{
		\begin{minipage}{0.93\linewidth}
			\textbf{Summary.} Different RFT failures induce different anomaly patterns, forming a structured empirical signature space rather than a single generic notion of abnormal training. While some faults are highly distinctive and strongly separated, others are subtler yet still remain distinguishable at a fine-grained level.
		\end{minipage}
	}
\end{center}

\section{RFT-FM: Automatic Failure Management for Reinforcement Fine-Tuning}

\begin{figure}[htbp]
	\centering
	\includegraphics[width=1\linewidth]{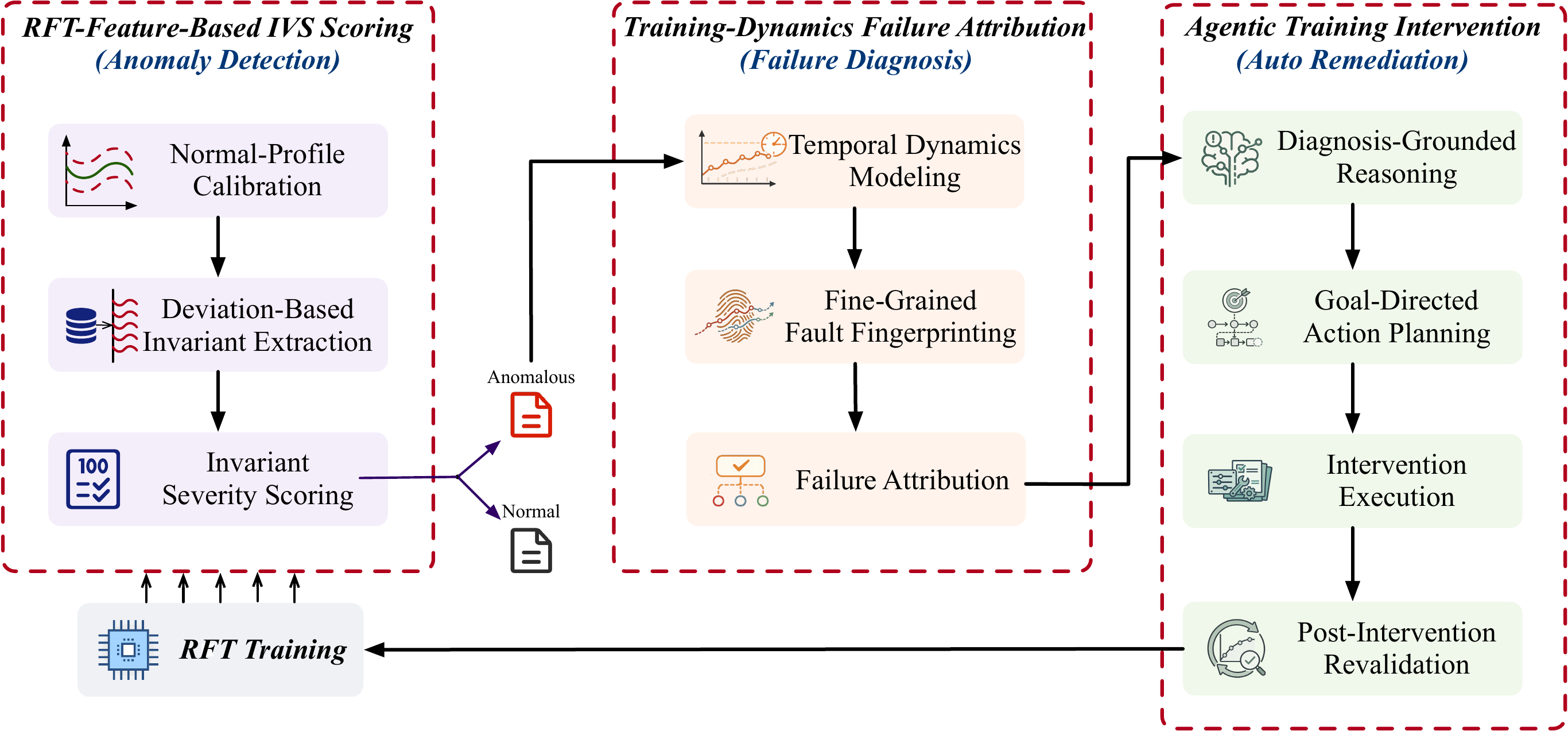}
	\caption{Architecture of RFT-FM}
	\label{fig: method-architecture}
\end{figure}

Building on the empirical findings that RFT training failures are both observable and distinguishable in training dynamics, we propose RFT-FM, an automatic failure management framework for reinforcement fine-tuning. As shown in Fig.~\ref{fig: method-architecture}, \textsc{RFT-FM} consists of three tightly coupled components. First, \textit{RFT-Feature-Based IVS Scoring} performs anomaly detection by calibrating healthy training profiles, extracting deviation-based invariant features, and aggregating them into a unified anomaly severity score. Second, \textit{Training-Dynamics Failure Attribution} performs failure diagnosis by modeling temporal training dynamics, constructing fine-grained fault fingerprints, and attributing detected anomalies to their most likely fault family and fault type. Third, \textit{Agentic Training Intervention} performs automatic remediation by reasoning over the diagnosed failure state, planning goal-directed corrective actions, executing intervention updates on the training configuration, and revalidating the resulting training dynamics in a closed loop. Together, these three components turn failure handling in RFT from manual expert inspection into a structured and automatable process.

\subsection{RFT-Feature-Based IVS Scoring}

We first address anomaly detection in reinforcement fine-tuning through \textit{RFT-Feature-Based IVS Scoring}. The goal of this module is to determine whether a training run deviates from healthy RFT dynamics, and if so, to quantify the severity of such deviation with a unified anomaly score. Unlike generic threshold-based monitoring, our detector is designed around RFT-specific training invariants and evaluates anomalies relative to the profile of normal training rather than in absolute isolation.

Formally, let a training run be represented as a multivariate trajectory \(X=\{x_t\}_{t=1}^{T}\), where each step-level observation \(x_t \in \mathbb{R}^{d}\) contains core RFT telemetry such as reward, KL divergence, entropy, return, response length, policy loss, and tool/environment interaction signals. The detection objective is to learn a severity scoring function that maps the trajectory \(X\) to a nonnegative anomaly score.

\paragraph{Normal-Profile Calibration.}
The first stage constructs a reference profile of healthy RFT training dynamics from normal runs. Let \(\mathcal{N}=\{X^{(1)},\dots,X^{(M)}\}\) denote the set of normal training runs. From these runs, we estimate a normal-profile calibration model \(\mathcal{P}_{\text{norm}}\), which summarizes the expected statistical range and dynamic tendency of healthy training signals. Intuitively, \(\mathcal{P}_{\text{norm}}\) defines what normal reward improvement, KL evolution, entropy decay, return stability, and generation behavior should look like under non-faulty RFT. This calibration is critical because anomaly detection in RFT is inherently relative: many signals are not anomalous in absolute magnitude, but become anomalous once they deviate from the characteristic profile of healthy training.

\paragraph{Deviation-Based Invariant Extraction.}
Given a new run \(X\), we do not directly score raw telemetry values. Instead, we transform them into a set of structured invariant deviations with respect to \(\mathcal{P}_{\text{norm}}\). Specifically, we define a collection of RFT-specific invariants \(\mathcal{I}=\{I_1,I_2,I_3,I_4,I_5\}\), corresponding to reward trajectory consistency, KL dynamics, entropy profile, return stability, and generation quality, respectively. Each invariant is evaluated through a set of derived statistics over the trajectory and converted into a deviation score relative to the calibrated normal profile. We denote the resulting invariant-deviation representation by

\begin{equation}
	z(X)=\left[z_1(X),z_2(X),\dots,z_K(X)\right],
	\label{eq:ivs-deviation-vector}
\end{equation}

In Eq.~\ref{eq:ivs-deviation-vector}, each component \(z_k(X)\) measures how strongly one invariant-related statistic departs from healthy behavior. This stage therefore maps raw RFT telemetry into a structured anomaly representation, so that anomaly characterization is performed in an invariant-aware and training-dynamics-aware space rather than directly in the raw signal space.

\paragraph{Invariant Severity Scoring.}
The final stage aggregates invariant deviations into a unified anomaly severity score. Let \(\phi(X)=\left[\phi_1(X),\phi_2(X),\dots,\phi_m(X)\right]\) denote the invariant-level deviation scores derived from \(z(X)\). We define the overall IVS score as

\begin{equation}
	S(X)=g(\phi(X)),
	\label{eq:ivs-score}
\end{equation}

Here, \(g(\cdot)\) is an aggregation function that combines multiple invariant violations into a single scalar severity measure. In practice, this aggregation emphasizes persistent and structured deviations across invariant dimensions rather than isolated fluctuations in a single signal. A run is then flagged as anomalous if \(S(X)>\tau\), where \(\tau\) is determined from the calibrated normal profile. Eq.~\ref{eq:ivs-score} defines the unified anomaly severity function, which is subsequently converted into a binary anomaly decision through thresholding.

Overall, this module implements anomaly detection as a three-stage process: \textit{Normal-Profile Calibration} defines the healthy reference regime, \textit{Deviation-Based Invariant Extraction} converts raw RFT dynamics into structured invariant deviations, and \textit{Invariant Severity Scoring} aggregates these deviations into a unified anomaly score. This design allows \textsc{RFT-FM} to detect RFT failures not merely as outliers in isolated metrics, but as structured departures from healthy reinforcement fine-tuning dynamics.

\subsection{Training-Dynamics Failure Attribution}

Given a detected anomalous run, the next goal of \textsc{RFT-FM} is not merely to confirm that training is abnormal, but to determine which failure mechanism is most likely responsible for the observed deviation. To this end, we design \textit{Training-Dynamics Failure Attribution}, a diagnosis module that maps anomalous training trajectories to their underlying fault family and fine-grained fault type. Unlike generic multi-class classification over static features, this module is centered on the temporal structure and cross-signal organization of RFT anomalies.

Formally, let \(X=\{x_t\}_{t=1}^{T}\) denote an anomalous training run that has already been identified by the detection module. The attribution objective is to learn a mapping from the anomalous trajectory \(X\) to a fault label \(y\), where \(y\) may correspond either to a fault family or to a fine-grained fault type. Instead of operating directly on raw telemetry, the attribution process first constructs a temporal representation of how the anomaly unfolds, then transforms this representation into a discriminative fault fingerprint, and finally performs structured fault attribution.

\paragraph{Temporal Dynamics Modeling.}
The first stage models the temporal evolution of the anomalous run. While detection mainly focuses on whether the run departs from healthy behavior, attribution requires a richer description of \emph{how} the deviation develops over time. For example, two faults may both exhibit degraded reward, but differ substantially in whether the degradation is abrupt or gradual, whether it co-occurs with KL explosion or entropy collapse, and whether generation behavior remains stable or deteriorates. We therefore summarize the run as a temporal dynamics representation:

\begin{equation}
	h(X)=\Psi\!\left(\{x_t\}_{t=1}^{T}\right),
	\label{eq:tdm-representation}
\end{equation}

Equation~\ref{eq:tdm-representation} defines \(h(X)\) as a temporal representation of the multivariate training trajectory, where \(\Psi(\cdot)\) denotes the temporal modeling operator. The resulting representation preserves dynamic information such as trend, fluctuation, relative timing, and cross-step consistency, allowing the diagnosis module to distinguish fault processes that may appear similar under static summaries.

\paragraph{Fine-Grained Fault Fingerprinting.}
Based on the temporal representation \(h(X)\), the second stage constructs a fine-grained fault fingerprint that captures the characteristic anomaly pattern of the run. Intuitively, this fingerprint is more discriminative than the severity-oriented representation used for detection: instead of only measuring how abnormal the run is, it describes which dimensions of reward, optimization, entropy, return, generation, and interaction dynamics are jointly affected and in what form. We denote the resulting fault fingerprint by:

\begin{equation}
	f(X)=\Phi\!\left(h(X)\right),
	\label{eq:fault-fingerprint}
\end{equation}

Equation~\ref{eq:fault-fingerprint} defines \(f(X)\) as a structured fault-level signature derived from temporal dynamics, where \(\Phi(\cdot)\) maps the temporal representation into a discriminative fingerprint space. This fingerprint is intended to preserve fine-grained distinctions among different RFT failures, including both coarse family-level differences and more subtle within-family variations. Such a design is important because many RFT anomalies are distinguishable not by a single dominant metric, but by the joint configuration of multiple training signals.

\paragraph{Failure Attribution.}
The final stage maps the fault fingerprint to its most likely diagnostic label. Let \(\mathcal{Y}_{\mathrm{fam}}\) and \(\mathcal{Y}_{\mathrm{type}}\) denote the family-level and type-level label spaces, respectively. The attribution module performs prediction in this structured fault space and outputs either a family-level diagnosis or a fine-grained fault-type diagnosis depending on the granularity required by downstream remediation. Concretely, attribution is performed as \( \hat{y}=A(f(X)) \), where \(A(\cdot)\) denotes the attribution function over the fingerprint representation. In this way, diagnosis is formulated not as a direct classification of raw telemetry, but as a mapping from structured anomaly fingerprints to interpretable fault labels.

Overall, this module performs failure diagnosis as a three-stage process: \textit{Temporal Dynamics Modeling} captures how the anomaly unfolds over training, \textit{Fine-Grained Fault Fingerprinting} converts this evolution into a discriminative fault signature, and \textit{Failure Attribution} maps the resulting signature to its most likely fault family and fault type. This design enables \textsc{RFT-FM} to move beyond anomaly detection and provide structured diagnostic explanations for RFT failures.

\subsection{Agentic Training Intervention}

Given the diagnosed fault family and fine-grained fault type, the final goal of \textsc{RFT-FM} is to move beyond passive monitoring and provide actionable remediation for anomalous RFT runs. To this end, we design \textit{Agentic Training Intervention}, a closed-loop intervention module that treats remediation not as a one-shot parameter suggestion task, but as a diagnosis-grounded and goal-directed decision process. Instead of directly emitting a static fix, this module reasons over the diagnosed failure state, plans corrective actions under training constraints, executes the intervention on the training configuration, and revalidates the resulting training dynamics.

Formally, let \(X\) denote a detected anomalous run, let \(\hat{y}\) denote its attributed fault label, and let \(c\) denote the current training configuration. The intervention objective is to produce an updated configuration \(\tilde{c}\) that mitigates the anomaly while preserving training feasibility. We formulate this process as a sequential agentic mapping from diagnosed training state to intervention outcome.

\paragraph{Diagnosis-Grounded Reasoning.}
The first stage interprets the diagnosed fault state and establishes the remediation objective. Rather than operating directly on raw telemetry alone, the intervention module conditions on the attributed fault label, anomaly severity, training-dynamics fingerprint, and current configuration. This stage identifies the likely failure mechanism and determines which aspects of training behavior should be restored, such as reward improvement, KL stabilization, entropy recovery, or generation normalization. In this way, remediation is grounded in structured diagnosis rather than ad hoc parameter adjustment.

\paragraph{Goal-Directed Action Planning.}
Based on the diagnosed state, the second stage plans a constrained corrective action over repairable training parameters. Let \(s(X,\hat{y},c)\) denote the structured intervention state induced by the anomalous run, its attributed fault label, and the current configuration. The action planner then produces an intervention action:

\begin{equation}
	a=\Pi\!\left(s(X,\hat{y},c)\right),
	\label{eq:intervention-action}
\end{equation}

In which, \(\Pi(\cdot)\) denotes the planning policy of the intervention agent. Equation~\ref{eq:intervention-action} defines the intervention action \(a\) as a diagnosis-conditioned corrective decision over controllable training parameters. In practice, the planner favors minimal but targeted changes, so that remediation remains interpretable and avoids unnecessarily large perturbations to the original training process.

\paragraph{Intervention Execution.}
The third stage executes the planned action by updating the current training configuration. Let \(\Gamma(\cdot)\) denote the execution operator that applies the planned action to the original configuration. The updated configuration is written as:

\begin{equation}
	\tilde{c}=\Gamma(c,a),
	\label{eq:updated-config}
\end{equation}

In which, \(\tilde{c}\) is the repaired configuration used to resume or restart training. Equation~\ref{eq:updated-config} makes explicit that intervention is not merely a textual recommendation, but an executable configuration update over the underlying training process. This stage therefore connects diagnosis to concrete system-level action.

\paragraph{Post-Intervention Revalidation.}
The final stage revalidates the intervention outcome by rerunning training under the updated configuration and observing the resulting dynamics. Let \(\tilde{X}\) denote the post-intervention training trajectory induced by \(\tilde{c}\). The intervention is then assessed through the updated anomaly severity \(S(\tilde{X})\), together with qualitative recovery of fault-relevant signals such as reward, KL, entropy, response length, and return stability. This stage closes the loop between intervention and monitoring: if anomaly severity decreases and the recovered dynamics move closer to the healthy regime, the intervention is considered mitigating; otherwise, the run remains unmitigated and may require further intervention.

Overall, this module performs remediation as a four-stage process: \textit{Diagnosis-Grounded Reasoning} interprets the failure state, \textit{Goal-Directed Action Planning} determines a constrained corrective action, \textit{Intervention Execution} applies the resulting update to the training configuration, and \textit{Post-Intervention Revalidation} evaluates whether the intervention has mitigated the anomaly. This design enables \textsc{RFT-FM} to extend failure management from anomaly observation and diagnosis to closed-loop, diagnosis-conditioned automatic remediation.

\section{Evaluation}

In this section, we evaluate RFT-FM from five perspectives. We first examine its effectiveness on anomaly detection, and then evaluate its ability to attribute detected anomalies to their underlying fault families and fine-grained fault types. Next, we conduct ablation studies and hyperparameter analyses to understand the contribution of core modules and the sensitivity of the framework to temporal horizon. Finally, we present a preliminary study on LLM-based automatic remediation to assess whether diagnosis-conditioned intervention can mitigate RFT anomaly severity.

\subsection{Experimental Setup}

We evaluate RFT-FM on RFT-FaultBench. Unless otherwise stated, RFT-FM uses a temporal horizon of 20 training steps, 5-fold cross-validation for both anomaly detection and failure diagnosis, and an IVS threshold coefficient of \(k=2.0\) for anomaly scoring. For automatic remediation, the agentic intervention module is instantiated with the qwen-plus model.

All experiments are conducted on a local compute server equipped with 160 Intel(R) Xeon(R) CPU cores, 1.8\,TiB RAM, and 8 NVIDIA H20 GPUs.

\subsection{Metrics}

For anomaly detection, we report precision, recall, and F1 score under both easy and hard settings. For failure diagnosis, we report macro-precision, macro-recall, and macro-F1 at the fault-type level.

For automatic remediation, we report two mitigation-oriented metrics. Let \(S(X)\) denote the anomaly severity score of an original faulty run \(X\), and let \(S(\tilde{X})\) denote the anomaly severity score after intervention. We first define an indicator function as Eq.~\ref{eq:indicator}, which equals 1 if the intervention reduces anomaly severity and 0 otherwise.

\begin{equation}
	\mathbb{I}\!\left(S(\tilde{X}) < S(X)\right),
	\label{eq:indicator}
\end{equation}

\textit{Mitigation Rate} measures the proportion of runs whose post-intervention severity is lower than that of the original faulty run. Formally, given a set of remediation cases \(\mathcal{D}_{\mathrm{rem}}\), Mitigation Rate is defined as Eq.~\ref{eq:mitigation-rate}.

\begin{equation}
	\mathrm{MR}
	=
	\frac{1}{|\mathcal{D}_{\mathrm{rem}}|}
	\sum_{X \in \mathcal{D}_{\mathrm{rem}}}
	\mathbb{I}\!\left(S(\tilde{X}) < S(X)\right).
	\label{eq:mitigation-rate}
\end{equation}

\textit{Median Severity Change} measures the median relative change in anomaly severity after intervention. For each run \(X\), we define its relative severity change as Eq.~\ref{eq:relative-severity-change}.

\begin{equation}
	\Delta(X) = \frac{S(X)-S(\tilde{X})}{S(X)} \times 100\%.
	\label{eq:relative-severity-change}
\end{equation}

A positive value of \(\Delta(X)\) indicates mitigation, while a negative value indicates deterioration. The overall Median Severity Change is then defined as Eq.~\ref{eq:median-severity-change}.

\begin{equation}
	\mathrm{MSC}
	=
	\mathrm{median}_{X \in \mathcal{D}_{\mathrm{rem}}}
	\left(
	\frac{S(X)-S(\tilde{X})}{S(X)} \times 100\%
	\right).
	\label{eq:median-severity-change}
\end{equation}

According to Eq.~\ref{eq:mitigation-rate}, Mitigation Rate captures how often the intervention moves the faulty run in the correct direction, while Eq.~\ref{eq:median-severity-change} measures the typical magnitude of such severity change across remediation cases.

\subsection{Compared Approaches}

We compare \textsc{RFT-FM} with representative baselines for both anomaly detection and failure diagnosis.

\paragraph{Anomaly Detection Baselines.}
For anomaly detection, we consider two groups of baselines. The first group includes classical run-level anomaly detectors, namely \textbf{IF} (Isolation Forest) and \textbf{LOF} (Local Outlier Factor), which operate on statistical features extracted from step-level RFT telemetry. Concretely, each training run is summarized by the mean, standard deviation, minimum, maximum, range, first-step value, last-step value, and linear trend of each metric, and anomaly detection is then performed in this feature space.

The second group includes sequence-based time-series anomaly detection methods, namely \textbf{TranAD}~\cite{tuli2022tranad}, \textbf{OmniAnomaly}~\cite{su2019robust}, and \textbf{AT} (Anomaly Transformer)~\cite{xuanomaly}. Since these methods are originally designed for generic multivariate time-series anomaly detection rather than RFT training analysis, we adapt them to the run-level RFT setting as follows: each training run is treated as a multivariate telemetry sequence over train steps; the detector is trained only on normal runs; per-step anomaly scores are first computed on the sequence and then aggregated into a run-level anomaly score for final evaluation. This setting provides a stronger comparison against modern sequence-based anomaly detectors while keeping the evaluation protocol consistent with RFT-FM.

\paragraph{Failure Diagnosis Baselines.}
For failure diagnosis, we compare RFT-FM with both generic classification baselines and RCA-inspired structured baselines. The generic baselines include \textbf{KNN} and \textbf{SVM}, which directly predict fault labels from the same statistical training-dynamics features described above. These methods test whether off-the-shelf classifiers are sufficient for distinguishing fine-grained RFT failures.

We further include three state-of-art baselines: \textbf{RUN}~\cite{lin2024root}, \textbf{CausalRCA}~\cite{xin2023causalrca}, and \textbf{CIRCA}~\cite{jiang2025circa}. Since these methods are originally proposed for root-cause analysis in microservice or system telemetry rather than RFT fault diagnosis, we adapt them to our setting by constructing interaction patterns over RFT telemetry signals. Specifically, we first derive a metric-interaction graph from the multivariate training trajectory using lagged dependency estimates, and then apply RCA-inspired feature construction for fault attribution. \textbf{RUN} is adapted into a graph-centrality-based attribution baseline, \textbf{CausalRCA} is adapted into a causal-strength prototype attribution baseline, and \textbf{CIRCA} is adapted into a graph-structural attribution baseline. These adaptations allow us to test whether existing structured RCA methods can transfer to the RFT failure diagnosis setting.

\begin{table*}[!t]
	\centering
	\setlength{\tabcolsep}{3.2pt}
	\renewcommand{\arraystretch}{1.15}
	\caption{Per-family anomaly detection results (\%).}
	\label{tab:detection-family-results}
	\begin{tabular}{cc|ccc|ccc|ccc|ccc|ccc|ccc}
		\toprule
		\multicolumn{2}{c|}{\multirow{2}{*}{\textbf{Dataset}}}
		& \multicolumn{3}{c|}{\textbf{IF}}
		& \multicolumn{3}{c|}{\textbf{LOF}}
		& \multicolumn{3}{c|}{\textbf{TranAD}}
		& \multicolumn{3}{c|}{\textbf{OmniAnomaly}}
		& \multicolumn{3}{c|}{\textbf{AT}}
		& \multicolumn{3}{c}{\textbf{RFT-FM (\textit{Ours})}} \\
		\cmidrule(lr){3-5}
		\cmidrule(lr){6-8}
		\cmidrule(lr){9-11}
		\cmidrule(lr){12-14}
		\cmidrule(lr){15-17}
		\cmidrule(lr){18-20}
		\multicolumn{2}{c|}{}
		& \textit{P} & \textit{R} & \textit{F1}
		& \textit{P} & \textit{R} & \textit{F1}
		& \textit{P} & \textit{R} & \textit{F1}
		& \textit{P} & \textit{R} & \textit{F1}
		& \textit{P} & \textit{R} & \textit{F1}
		& \textit{P} & \textit{R} & \textit{F1} \\
		\midrule
		
		\multirow{2}{*}{RF} & \cellcolor{acccolor}Easy
		& \cellcolor{acccolor}92.35 & \cellcolor{acccolor}66.67 & \cellcolor{acccolor}75.53
		& \cellcolor{acccolor}22.22 & \cellcolor{acccolor}3.33 & \cellcolor{acccolor}5.80
		& \cellcolor{acccolor}100.00 & \cellcolor{acccolor}90.00 & \cellcolor{acccolor}\textbf{94.59}
		& \cellcolor{acccolor}100.00 & \cellcolor{acccolor}66.67 & \cellcolor{acccolor}72.66
		& \cellcolor{acccolor}100.00 & \cellcolor{acccolor}66.67 & \cellcolor{acccolor}70.97
		& \cellcolor{acccolor}94.00 & \cellcolor{acccolor}81.67 & \cellcolor{acccolor}\underline{86.59} \\
		
		& \cellcolor{latcolor}Hard
		& \cellcolor{latcolor}16.67 & \cellcolor{latcolor}1.19 & \cellcolor{latcolor}2.22
		& \cellcolor{latcolor}0.00 & \cellcolor{latcolor}0.00 & \cellcolor{latcolor}0.00
		& \cellcolor{latcolor}100.00 & \cellcolor{latcolor}63.10 & \cellcolor{latcolor}\textbf{77.28}
		& \cellcolor{latcolor}66.67 & \cellcolor{latcolor}25.00 & \cellcolor{latcolor}35.56
		& \cellcolor{latcolor}100.00 & \cellcolor{latcolor}58.33 & \cellcolor{latcolor}72.20
		& \cellcolor{latcolor}94.51 & \cellcolor{latcolor}63.10 & \cellcolor{latcolor}\underline{75.24} \\
		
		\midrule
		
		\multirow{2}{*}{PG} & \cellcolor{acccolor}Easy
		& \cellcolor{acccolor}71.23 & \cellcolor{acccolor}71.25 & \cellcolor{acccolor}71.15
		& \cellcolor{acccolor}47.62 & \cellcolor{acccolor}50.00 & \cellcolor{acccolor}48.78
		& \cellcolor{acccolor}75.00 & \cellcolor{acccolor}75.00 & \cellcolor{acccolor}75.00
		& \cellcolor{acccolor}75.00 & \cellcolor{acccolor}75.00 & \cellcolor{acccolor}75.00
		& \cellcolor{acccolor}100.00 & \cellcolor{acccolor}81.25 & \cellcolor{acccolor}\textbf{85.00}
		& \cellcolor{acccolor}83.93 & \cellcolor{acccolor}76.25 & \cellcolor{acccolor}\underline{75.44} \\
		
		& \cellcolor{latcolor}Hard
		& \cellcolor{latcolor}0.00 & \cellcolor{latcolor}0.00 & \cellcolor{latcolor}0.00
		& \cellcolor{latcolor}50.00 & \cellcolor{latcolor}17.85 & \cellcolor{latcolor}26.31
		& \cellcolor{latcolor}100.00 & \cellcolor{latcolor}57.15 & \cellcolor{latcolor}71.97
		& \cellcolor{latcolor}75.00 & \cellcolor{latcolor}30.35 & \cellcolor{latcolor}42.98
		& \cellcolor{latcolor}100.00 & \cellcolor{latcolor}63.40 & \cellcolor{latcolor}\textbf{77.31}
		& \cellcolor{latcolor}94.08 & \cellcolor{latcolor}60.71 & \cellcolor{latcolor}\underline{72.80} \\
		
		\midrule
		
		\multirow{2}{*}{OD} & \cellcolor{acccolor}Easy
		& \cellcolor{acccolor}93.72 & \cellcolor{acccolor}81.67 & \cellcolor{acccolor}85.69
		& \cellcolor{acccolor}31.48 & \cellcolor{acccolor}28.33 & \cellcolor{acccolor}29.82
		& \cellcolor{acccolor}100.00 & \cellcolor{acccolor}100.00 & \cellcolor{acccolor}\textbf{100.00}
		& \cellcolor{acccolor}66.67 & \cellcolor{acccolor}66.67 & \cellcolor{acccolor}66.67
		& \cellcolor{acccolor}100.00 & \cellcolor{acccolor}75.00 & \cellcolor{acccolor}80.00
		& \cellcolor{acccolor}95.24 & \cellcolor{acccolor}100.00 & \cellcolor{acccolor}\underline{97.56} \\
		
		& \cellcolor{latcolor}Hard
		& \cellcolor{latcolor}26.67 & \cellcolor{latcolor}4.76 & \cellcolor{latcolor}8.08
		& \cellcolor{latcolor}0.00 & \cellcolor{latcolor}0.00 & \cellcolor{latcolor}0.00
		& \cellcolor{latcolor}100.00 & \cellcolor{latcolor}61.91 & \cellcolor{latcolor}\textbf{76.29}
		& \cellcolor{latcolor}66.67 & \cellcolor{latcolor}19.05 & \cellcolor{latcolor}29.63
		& \cellcolor{latcolor}100.00 & \cellcolor{latcolor}59.53 & \cellcolor{latcolor}74.40
		& \cellcolor{latcolor}94.65 & \cellcolor{latcolor}64.29 & \cellcolor{latcolor}\underline{76.26} \\
		
		\midrule
		
		\multirow{2}{*}{CA} & \cellcolor{acccolor}Easy
		& \cellcolor{acccolor}53.33 & \cellcolor{acccolor}13.33 & \cellcolor{acccolor}21.33
		& \cellcolor{acccolor}27.78 & \cellcolor{acccolor}8.33 & \cellcolor{acccolor}12.82
		& \cellcolor{acccolor}66.67 & \cellcolor{acccolor}63.33 & \cellcolor{acccolor}\underline{64.91}
		& \cellcolor{acccolor}66.67 & \cellcolor{acccolor}61.67 & \cellcolor{acccolor}64.06
		& \cellcolor{acccolor}100.00 & \cellcolor{acccolor}63.33 & \cellcolor{acccolor}\textbf{72.93}
		& \cellcolor{acccolor}61.97 & \cellcolor{acccolor}48.33 & \cellcolor{acccolor}53.17 \\
		
		& \cellcolor{latcolor}Hard
		& \cellcolor{latcolor}50.00 & \cellcolor{latcolor}7.14 & \cellcolor{latcolor}12.50
		& \cellcolor{latcolor}0.00 & \cellcolor{latcolor}0.00 & \cellcolor{latcolor}0.00
		& \cellcolor{latcolor}100.00 & \cellcolor{latcolor}52.38 & \cellcolor{latcolor}\underline{68.74}
		& \cellcolor{latcolor}33.33 & \cellcolor{latcolor}4.76 & \cellcolor{latcolor}8.33
		& \cellcolor{latcolor}100.00 & \cellcolor{latcolor}50.00 & \cellcolor{latcolor}66.67
		& \cellcolor{latcolor}94.03 & \cellcolor{latcolor}57.14 & \cellcolor{latcolor}\textbf{70.82} \\
		
		\midrule
		
		\multirow{2}{*}{TE} & \cellcolor{acccolor}Easy
		& \cellcolor{acccolor}94.26 & \cellcolor{acccolor}86.67 & \cellcolor{acccolor}89.28
		& \cellcolor{acccolor}91.90 & \cellcolor{acccolor}73.33 & \cellcolor{acccolor}78.10
		& \cellcolor{acccolor}100.00 & \cellcolor{acccolor}96.67 & \cellcolor{acccolor}\textbf{98.25}
		& \cellcolor{acccolor}100.00 & \cellcolor{acccolor}91.67 & \cellcolor{acccolor}\underline{95.24}
		& \cellcolor{acccolor}100.00 & \cellcolor{acccolor}90.00 & \cellcolor{acccolor}94.38
		& \cellcolor{acccolor}94.45 & \cellcolor{acccolor}88.33 & \cellcolor{acccolor}90.53 \\
		
		& \cellcolor{latcolor}Hard
		& \cellcolor{latcolor}29.17 & \cellcolor{latcolor}8.33 & \cellcolor{latcolor}12.96
		& \cellcolor{latcolor}33.33 & \cellcolor{latcolor}11.90 & \cellcolor{latcolor}17.54
		& \cellcolor{latcolor}66.67 & \cellcolor{latcolor}40.48 & \cellcolor{latcolor}50.21
		& \cellcolor{latcolor}100.00 & \cellcolor{latcolor}32.14 & \cellcolor{latcolor}46.77
		& \cellcolor{latcolor}100.00 & \cellcolor{latcolor}42.86 & \cellcolor{latcolor}\underline{57.05}
		& \cellcolor{latcolor}89.73 & \cellcolor{latcolor}47.62 & \cellcolor{latcolor}\textbf{58.81} \\
		
		\midrule
		
		\multirow{2}{*}{\textbf{Avg.}} & \cellcolor{acccolor}\textbf{Easy}
		& \cellcolor{acccolor}81.98 & \cellcolor{acccolor}63.12 & \cellcolor{acccolor}68.60
		& \cellcolor{acccolor}44.20 & \cellcolor{acccolor}32.66 & \cellcolor{acccolor}35.06
		& \cellcolor{acccolor}88.33 & \cellcolor{acccolor}85.00 & \cellcolor{acccolor}\textbf{86.55}
		& \cellcolor{acccolor}81.67 & \cellcolor{acccolor}72.34 & \cellcolor{acccolor}74.73
		& \cellcolor{acccolor}100.00 & \cellcolor{acccolor}75.25 & \cellcolor{acccolor}\underline{80.66}
		& \cellcolor{acccolor}85.92 & \cellcolor{acccolor}78.92 & \cellcolor{acccolor}\underline{80.66} \\
		
		& \cellcolor{latcolor}\textbf{Hard}
		& \cellcolor{latcolor}24.50 & \cellcolor{latcolor}4.28 & \cellcolor{latcolor}7.15
		& \cellcolor{latcolor}16.67 & \cellcolor{latcolor}5.95 & \cellcolor{latcolor}8.77
		& \cellcolor{latcolor}93.33 & \cellcolor{latcolor}55.00 & \cellcolor{latcolor}68.90
		& \cellcolor{latcolor}68.33 & \cellcolor{latcolor}22.26 & \cellcolor{latcolor}32.65
		& \cellcolor{latcolor}100.00 & \cellcolor{latcolor}54.82 & \cellcolor{latcolor}\underline{69.53}
		& \cellcolor{latcolor}93.40 & \cellcolor{latcolor}58.57 & \cellcolor{latcolor}\textbf{70.75} \\
		
		\bottomrule
	\end{tabular}
\end{table*}

\subsection{Anomaly Detection}

We first evaluate the anomaly detection capability of RFT-FM, aiming to examine whether RFT failures can be reliably identified from training dynamics under both easy and hard settings. As shown in Table~\ref{tab:detection-family-results}, RFT-FM achieves consistently strong anomaly detection performance across most fault families under both easy and hard settings, while the behavior of competing baselines also reveals that RFT-FaultBench is neither trivial nor saturated.

Under the easy setting, several strong sequence-based baselines already obtain competitive results on highly visible failures such as RF, OD, and TE. For example, TranAD reaches 94.59\% F1 on RF, 100.00\% on OD, and 98.25\% on TE, while Anomaly Transformer also performs strongly on PG and CA. This shows that many RFT failures indeed induce clear anomaly structure in training dynamics. However, the benchmark is far from saturated. Even in the easy setting, different methods exhibit substantial variation across fault families, and some failures remain clearly more difficult than others. In particular, CA faults are still challenging for multiple baselines, and \textsc{RFT-FM} itself achieves only 53.17\% F1 on easy CA, much lower than its results on RF, OD, or TE. This confirms that observability in \textsc{RFT-FaultBench} is structured rather than uniform.

The harder setting further highlights the nontrivial nature of the benchmark. Most classical statistical baselines degrade sharply: IF and LOF collapse on nearly all families, and even strong deep detectors show much larger performance gaps across families. Although TranAD and Anomaly Transformer remain competitive on some hard faults, their robustness is still uneven. In contrast, RFT-FM maintains consistently strong family-wise performance under the hard setting, achieving F1 scores of 75.24\%, 72.80\%, 76.26\%, 70.82\%, and 58.81\% on RF, PG, OD, CA, and TE, respectively, with the best average hard-setting F1 of 70.75\%. These results indicate that RFT-FM better captures structured deviations specific to RFT training, rather than relying only on easily separable surface-level anomalies.

Overall, the detection results support two main conclusions. First, RFT-FaultBench is neither trivial nor saturated: it exhibits clear anomaly structure, yet still poses substantial challenges, especially under subtle fault settings. Second, RFT-FM shows strong anomaly detection capability across diverse RFT failure families and demonstrates better robustness than generic baselines in the harder setting.

\begin{table*}[htbp]
	\centering
	\setlength{\tabcolsep}{3.8pt}
	\renewcommand{\arraystretch}{1.15}
	\caption{Per-family failure diagnosis results (\%).}
	\label{tab:diagnosis-family-results}
	\begin{tabular}{cc|ccc|ccc|ccc|ccc|ccc|ccc}
		\toprule
		\multicolumn{2}{c|}{\multirow{2}{*}{\textbf{Dataset}}}
		& \multicolumn{3}{c|}{\textbf{KNN}}
		& \multicolumn{3}{c|}{\textbf{SVM}}
		& \multicolumn{3}{c|}{\textbf{RUN}}
		& \multicolumn{3}{c|}{\textbf{CausalRCA}}
		& \multicolumn{3}{c|}{\textbf{CIRCA}}
		& \multicolumn{3}{c}{\textbf{RFT-FM (\textit{Ours})}} \\
		\cmidrule(lr){3-5}
		\cmidrule(lr){6-8}
		\cmidrule(lr){9-11}
		\cmidrule(lr){12-14}
		\cmidrule(lr){15-17}
		\cmidrule(lr){18-20}
		\multicolumn{2}{c|}{}
		& \textit{P} & \textit{R} & \textit{F1}
		& \textit{P} & \textit{R} & \textit{F1}
		& \textit{P} & \textit{R} & \textit{F1}
		& \textit{P} & \textit{R} & \textit{F1}
		& \textit{P} & \textit{R} & \textit{F1}
		& \textit{P} & \textit{R} & \textit{F1} \\
		\midrule
		
		\multirow{2}{*}{RF} & \cellcolor{acccolor}Easy
		& \cellcolor{acccolor}90.93 & \cellcolor{acccolor}88.33 & \cellcolor{acccolor}89.56
		& \cellcolor{acccolor}93.94 & \cellcolor{acccolor}96.67 & \cellcolor{acccolor}\underline{95.24}
		& \cellcolor{acccolor}28.38 & \cellcolor{acccolor}35.00 & \cellcolor{acccolor}31.21
		& \cellcolor{acccolor}43.83 & \cellcolor{acccolor}63.33 & \cellcolor{acccolor}51.77
		& \cellcolor{acccolor}67.54 & \cellcolor{acccolor}61.67 & \cellcolor{acccolor}64.46
		& \cellcolor{acccolor}96.83 & \cellcolor{acccolor}95.00 & \cellcolor{acccolor}\textbf{95.67} \\
		
		& \cellcolor{latcolor}Hard
		& \cellcolor{latcolor}65.71 & \cellcolor{latcolor}52.38 & \cellcolor{latcolor}36.35
		& \cellcolor{latcolor}73.70 & \cellcolor{latcolor}27.38 & \cellcolor{latcolor}39.81
		& \cellcolor{latcolor}0.00 & \cellcolor{latcolor}0.00 & \cellcolor{latcolor}0.00
		& \cellcolor{latcolor}0.00 & \cellcolor{latcolor}0.00 & \cellcolor{latcolor}0.00
		& \cellcolor{latcolor}24.07 & \cellcolor{latcolor}4.76 & \cellcolor{latcolor}7.77
		& \cellcolor{latcolor}75.20 & \cellcolor{latcolor}40.47 & \cellcolor{latcolor}\textbf{52.59} \\
		
		\midrule
		
		\multirow{2}{*}{PG} & \cellcolor{acccolor}Easy
		& \cellcolor{acccolor}81.33 & \cellcolor{acccolor}91.25 & \cellcolor{acccolor}\underline{85.74}
		& \cellcolor{acccolor}69.62 & \cellcolor{acccolor}75.00 & \cellcolor{acccolor}71.05
		& \cellcolor{acccolor}27.95 & \cellcolor{acccolor}41.25 & \cellcolor{acccolor}31.51
		& \cellcolor{acccolor}44.97 & \cellcolor{acccolor}51.25 & \cellcolor{acccolor}46.95
		& \cellcolor{acccolor}52.04 & \cellcolor{acccolor}67.50 & \cellcolor{acccolor}56.74
		& \cellcolor{acccolor}85.20 & \cellcolor{acccolor}93.75 & \cellcolor{acccolor}\textbf{89.11} \\
		
		& \cellcolor{latcolor}Hard
		& \cellcolor{latcolor}63.19 & \cellcolor{latcolor}40.18 & \cellcolor{latcolor}\underline{45.19}
		& \cellcolor{latcolor}60.66 & \cellcolor{latcolor}39.29 & \cellcolor{latcolor}43.18
		& \cellcolor{latcolor}11.30 & \cellcolor{latcolor}8.04 & \cellcolor{latcolor}9.06
		& \cellcolor{latcolor}1.70 & \cellcolor{latcolor}2.68 & \cellcolor{latcolor}2.08
		& \cellcolor{latcolor}17.99 & \cellcolor{latcolor}11.61 & \cellcolor{latcolor}13.22
		& \cellcolor{latcolor}37.58 & \cellcolor{latcolor}50.00 & \cellcolor{latcolor}\textbf{42.56} \\
		
		\midrule
		
		\multirow{2}{*}{OD} & \cellcolor{acccolor}Easy
		& \cellcolor{acccolor}98.41 & \cellcolor{acccolor}95.00 & \cellcolor{acccolor}\textbf{96.48}
		& \cellcolor{acccolor}94.71 & \cellcolor{acccolor}93.33 & \cellcolor{acccolor}93.92
		& \cellcolor{acccolor}65.26 & \cellcolor{acccolor}70.00 & \cellcolor{acccolor}67.24
		& \cellcolor{acccolor}80.30 & \cellcolor{acccolor}71.67 & \cellcolor{acccolor}72.77
		& \cellcolor{acccolor}95.65 & \cellcolor{acccolor}81.67 & \cellcolor{acccolor}87.03
		& \cellcolor{acccolor}98.04 & \cellcolor{acccolor}93.33 & \cellcolor{acccolor}\underline{95.50} \\
		
		& \cellcolor{latcolor}Hard
		& \cellcolor{latcolor}96.97 & \cellcolor{latcolor}26.19 & \cellcolor{latcolor}40.24
		& \cellcolor{latcolor}96.97 & \cellcolor{latcolor}28.57 & \cellcolor{latcolor}43.67
		& \cellcolor{latcolor}2.64 & \cellcolor{latcolor}16.67 & \cellcolor{latcolor}4.55
		& \cellcolor{latcolor}15.46 & \cellcolor{latcolor}29.76 & \cellcolor{latcolor}7.99
		& \cellcolor{latcolor}83.07 & \cellcolor{latcolor}23.81 & \cellcolor{latcolor}36.95
		& \cellcolor{latcolor}78.16 & \cellcolor{latcolor}38.09 & \cellcolor{latcolor}\textbf{48.73} \\
		
		\midrule
		
		\multirow{2}{*}{CA} & \cellcolor{acccolor}Easy
		& \cellcolor{acccolor}82.91 & \cellcolor{acccolor}88.33 & \cellcolor{acccolor}84.12
		& \cellcolor{acccolor}72.72 & \cellcolor{acccolor}81.67 & \cellcolor{acccolor}72.36
		& \cellcolor{acccolor}17.30 & \cellcolor{acccolor}13.33 & \cellcolor{acccolor}14.19
		& \cellcolor{acccolor}21.72 & \cellcolor{acccolor}18.33 & \cellcolor{acccolor}19.44
		& \cellcolor{acccolor}53.24 & \cellcolor{acccolor}66.67 & \cellcolor{acccolor}58.75
		& \cellcolor{acccolor}82.64 & \cellcolor{acccolor}95.00 & \cellcolor{acccolor}\textbf{88.28} \\
		
		& \cellcolor{latcolor}Hard
		& \cellcolor{latcolor}11.11 & \cellcolor{latcolor}2.38 & \cellcolor{latcolor}3.92
		& \cellcolor{latcolor}5.88 & \cellcolor{latcolor}16.67 & \cellcolor{latcolor}8.62
		& \cellcolor{latcolor}3.12 & \cellcolor{latcolor}14.29 & \cellcolor{latcolor}5.13
		& \cellcolor{latcolor}0.00 & \cellcolor{latcolor}0.00 & \cellcolor{latcolor}0.00
		& \cellcolor{latcolor}9.64 & \cellcolor{latcolor}35.71 & \cellcolor{latcolor}14.38
		& \cellcolor{latcolor}57.51 & \cellcolor{latcolor}35.71 & \cellcolor{latcolor}\textbf{39.34} \\
		
		\midrule
		
		\multirow{2}{*}{TE} & \cellcolor{acccolor}Easy
		& \cellcolor{acccolor}100.00 & \cellcolor{acccolor}86.67 & \cellcolor{acccolor}\textbf{92.36}
		& \cellcolor{acccolor}66.67 & \cellcolor{acccolor}65.00 & \cellcolor{acccolor}65.81
		& \cellcolor{acccolor}30.26 & \cellcolor{acccolor}23.33 & \cellcolor{acccolor}24.93
		& \cellcolor{acccolor}39.26 & \cellcolor{acccolor}38.33 & \cellcolor{acccolor}37.97
		& \cellcolor{acccolor}55.83 & \cellcolor{acccolor}50.00 & \cellcolor{acccolor}52.59
		& \cellcolor{acccolor}97.22 & \cellcolor{acccolor}83.33 & \cellcolor{acccolor}\underline{88.73} \\
		
		& \cellcolor{latcolor}Hard
		& \cellcolor{latcolor}79.64 & \cellcolor{latcolor}45.24 & \cellcolor{latcolor}46.98
		& \cellcolor{latcolor}65.94 & \cellcolor{latcolor}41.67 & \cellcolor{latcolor}48.59
		& \cellcolor{latcolor}8.33 & \cellcolor{latcolor}1.19 & \cellcolor{latcolor}2.08
		& \cellcolor{latcolor}33.33 & \cellcolor{latcolor}1.19 & \cellcolor{latcolor}2.30
		& \cellcolor{latcolor}44.44 & \cellcolor{latcolor}11.90 & \cellcolor{latcolor}17.03
		& \cellcolor{latcolor}82.77 & \cellcolor{latcolor}63.10 & \cellcolor{latcolor}\textbf{67.57} \\
		
		\midrule
		
		\multirow{2}{*}{\textbf{Avg.}} & \cellcolor{acccolor}\textbf{Easy}
		& \cellcolor{acccolor}90.72 & \cellcolor{acccolor}89.92 & \cellcolor{acccolor}\underline{89.66}
		& \cellcolor{acccolor}79.53 & \cellcolor{acccolor}82.33 & \cellcolor{acccolor}79.68
		& \cellcolor{acccolor}33.83 & \cellcolor{acccolor}36.58 & \cellcolor{acccolor}33.82
		& \cellcolor{acccolor}46.02 & \cellcolor{acccolor}48.58 & \cellcolor{acccolor}45.78
		& \cellcolor{acccolor}64.86 & \cellcolor{acccolor}65.50 & \cellcolor{acccolor}63.91
		& \cellcolor{acccolor}91.99 & \cellcolor{acccolor}92.08 & \cellcolor{acccolor}\textbf{91.46} \\
		
		& \cellcolor{latcolor}\textbf{Hard}
		& \cellcolor{latcolor}63.32 & \cellcolor{latcolor}33.27 & \cellcolor{latcolor}34.54
		& \cellcolor{latcolor}60.63 & \cellcolor{latcolor}30.71 & \cellcolor{latcolor}36.78
		& \cellcolor{latcolor}5.08 & \cellcolor{latcolor}8.04 & \cellcolor{latcolor}4.16
		& \cellcolor{latcolor}10.10 & \cellcolor{latcolor}6.73 & \cellcolor{latcolor}2.47
		& \cellcolor{latcolor}35.84 & \cellcolor{latcolor}17.56 & \cellcolor{latcolor}17.87
		& \cellcolor{latcolor}66.24 & \cellcolor{latcolor}45.48 & \cellcolor{latcolor}\textbf{50.16} \\
		
		\bottomrule
	\end{tabular}
\end{table*}

\subsection{Failure Diagnosis}

We then evaluate the failure diagnosis capability of RFT-FM, aiming to examine whether detected anomalous runs can be further attributed to their underlying fault families and fine-grained fault types. As shown in Table~\ref{tab:diagnosis-family-results}, RFT-FM achieves the strongest overall diagnosis performance under both easy and hard settings, while the comparison with competing baselines further shows that \textsc{RFT-FaultBench} remains challenging even after anomalies have become detectable.

Under the easy setting, several generic classification baselines already achieve strong results on relatively distinctive fault families. For example, KNN and SVM perform competitively on RF, OD, and TE, and even outperform RFT-FM on easy OD and easy TE in terms of F1. This indicates that some RFT failures produce sufficiently stable and separable signatures once anomaly visibility is strong. However, the benchmark is not saturated even in this easier regime. Structured RCA-inspired baselines such as RUN, CausalRCA, and CIRCA remain substantially weaker than generic classifiers on most families, and diagnosis difficulty still varies greatly across fault types. In contrast, RFT-FM achieves the best average easy-setting F1 of 91.46\%, outperforming all structured baselines and remaining competitive even against the strongest generic classifiers.

The harder diagnosis setting reveals a much sharper performance gap. All compared baselines degrade substantially, especially the RCA-inspired methods, many of which collapse to near-zero F1 on multiple fault families. Even strong generic classifiers such as KNN and SVM suffer large drops under subtle failure settings, with average hard-setting F1 scores of 34.54\% and 36.78\%, respectively. By comparison, RFT-FM maintains the best average hard-setting F1 of 50.16\%, achieving 52.59\%, 42.56\%, 48.73\%, 39.34\%, and 67.57\% on RF, PG, OD, CA, and TE, respectively. These results show that fine-grained fault attribution under subtle anomalies is substantially harder than anomaly detection alone, and that effective diagnosis requires modeling structured temporal fault fingerprints rather than relying only on generic static classification signals.

Overall, the diagnosis results support two conclusions. First, RFT-FaultBench is not only a detection benchmark but also a challenging diagnosis benchmark: while many failures remain distinguishable under clear anomaly settings, subtle failures still pose substantial difficulty for both generic classifiers and RCA-inspired methods. Second, RFT-FM shows strong capability in diagnosing RFT failures and provides a more robust attribution mechanism under hard settings than competing approaches.

\subsection{Auto Remediation}

We finally evaluate the automatic remediation capability of RFT-FM, aiming to examine whether diagnosis-conditioned agentic intervention can mitigate anomaly severity after a fault has been detected and attributed. Since automatic remediation remains substantially more challenging than anomaly detection and diagnosis, we treat this part as a preliminary closed-loop evaluation rather than a fully saturated benchmark task.

\begin{table}[htbp]
	\centering
	\setlength{\tabcolsep}{15pt}
	\caption{Mitigation Results of Automatic Remediation}
	\label{tab:llm_remediation_family}
	\begin{tabular}{lcc}
		\toprule
		Family & Mitigation Rate & Median Severity Change \\
		\midrule
		RF & 40.00\% & -63.26\% \\
		PG & 60.00\% & 17.56\% \\
		OD & 26.67\% & -26.75\% \\
		CA & 26.67\% & -38.10\% \\
		TE & 73.33\% & 13.13\% \\
		\midrule
		Overall & 46.25\%  & -5.84\%  \\
		\bottomrule
	\end{tabular}
\end{table}

Table~\ref{tab:llm_remediation_family} reports the family-level mitigation results. Overall, \textsc{RFT-FM} achieves a mitigation rate of 46.25\%, showing that nearly half of the remediation cases can be moved in the correct direction after intervention. However, the overall median severity change is -5.84\%, indicating that automatic remediation remains unstable and that unsuccessful interventions can still substantially worsen training dynamics. This result suggests that closing the loop from diagnosis to intervention is feasible, but still far from solved.

The remediation effectiveness is also highly heterogeneous across fault families. RFT-FM performs best on PG and TE faults, achieving mitigation rates of 60.00\% and 73.33\%, with positive median severity changes of 17.56\% and 13.13\%, respectively. These results indicate that policy-generation and tool/environment faults are relatively more amenable to diagnosis-conditioned intervention. By contrast, RF, OD, and CA remain much harder: although some cases can still be mitigated, their median severity changes are negative, showing that one-shot intervention is often insufficient and may even destabilize training further.

\begin{figure}[htbp]
	\centering
	\includegraphics[width=1\linewidth]{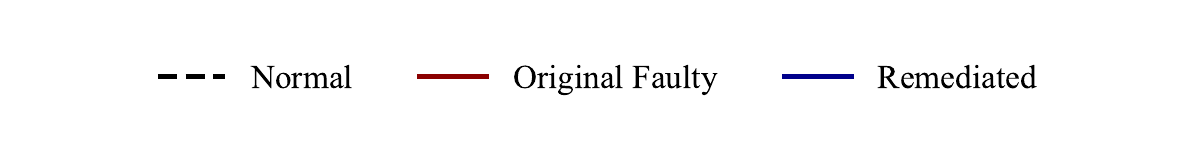}
	\vspace{0.4em}
	\subfigure[Reward]{
		\begin{minipage}{0.46\linewidth}
			\centering
			\includegraphics[width=\linewidth]{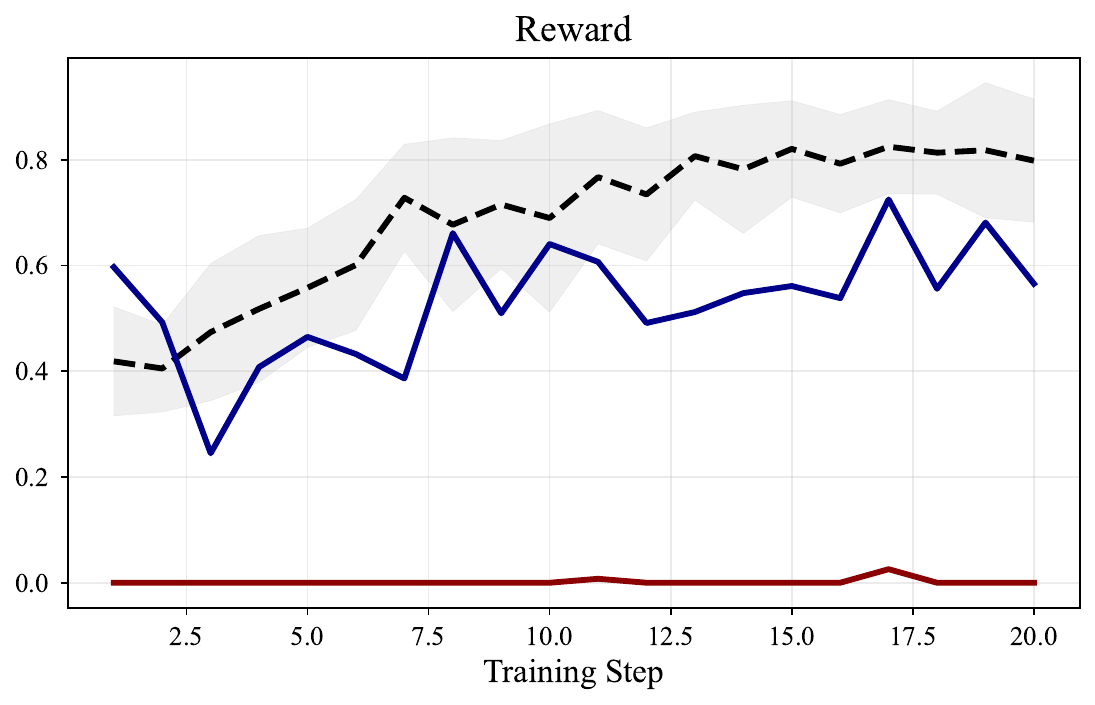}
			\label{fig:remediation-case-reward}
		\end{minipage}
	}
	\subfigure[KL Divergence]{
		\begin{minipage}{0.46\linewidth}
			\centering
			\includegraphics[width=\linewidth]{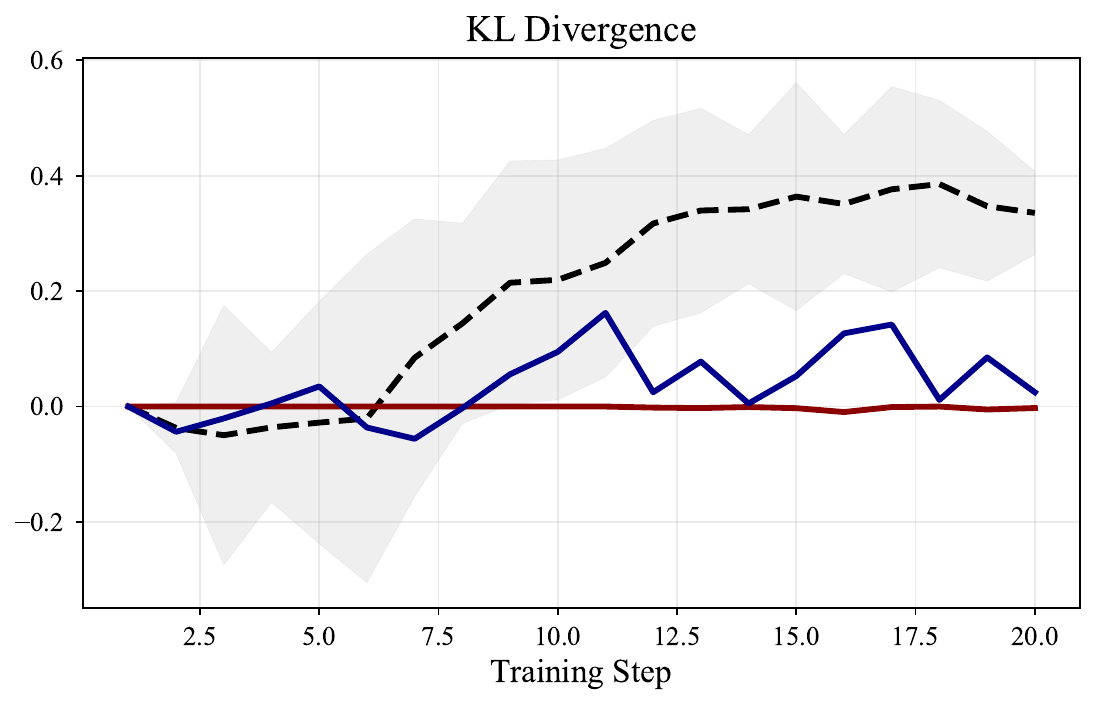}
			\label{fig:remediation-case-kl}
		\end{minipage}
	}
	\subfigure[Entropy]{
		\begin{minipage}{0.46\linewidth}
			\centering
			\includegraphics[width=\linewidth]{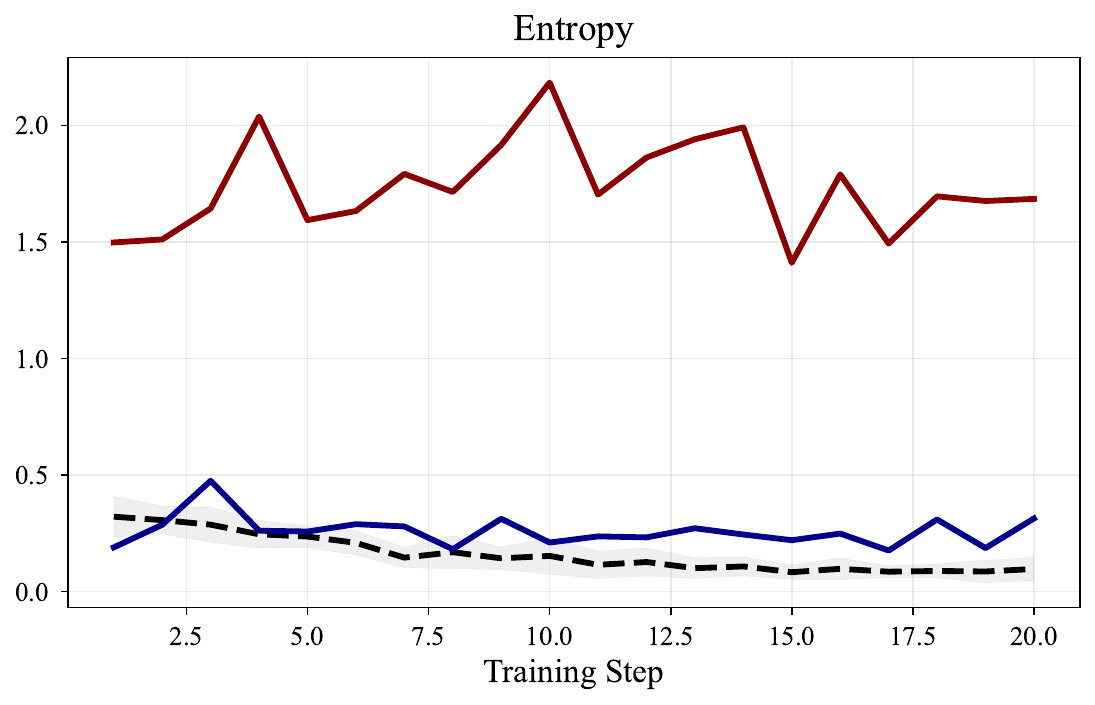}
			\label{fig:remediation-case-entropy}
		\end{minipage}
	}
	\subfigure[Response Length]{
		\begin{minipage}{0.46\linewidth}
			\centering
			\includegraphics[width=\linewidth]{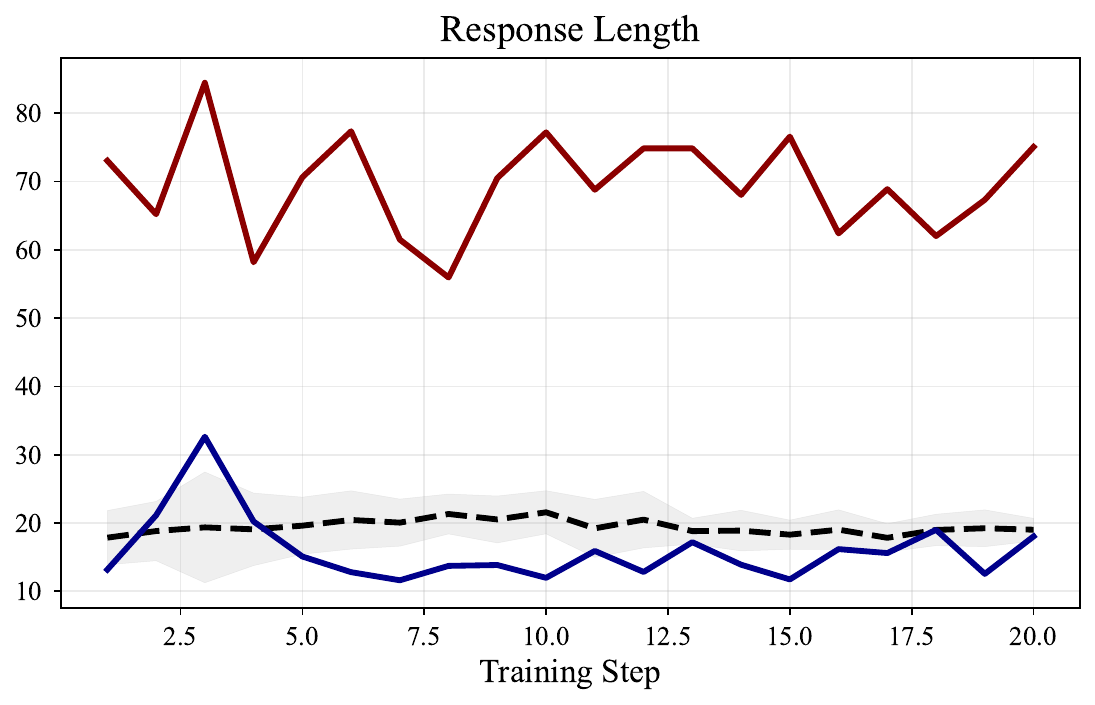}
			\label{fig:remediation-case-length}
		\end{minipage}
	}
	
	\caption{Case Study of RFT-FM Remediation (TE-2)}
	\label{fig:remediation-case}
\end{figure}

Figure~\ref{fig:remediation-case} provides a representative TE-2 case study. Compared with the original faulty run, the remediated run moves substantially closer to the normal training regime across multiple telemetry dimensions, including reward, KL divergence, entropy, and response length. This example illustrates what successful mitigation looks like in training dynamics: remediation does not simply reduce a scalar anomaly score, but partially restores the underlying trajectory structure toward healthy RFT behavior.

Overall, the remediation results support two conclusions. First, RFT-FM already provides a meaningful preliminary capability for diagnosis-conditioned automatic intervention, especially for fault families with clearer and more actionable failure signatures. Second, automatic remediation is considerably more difficult than detection and diagnosis, and robust closed-loop intervention for subtle or highly entangled RFT failures remains an important open problem.

\subsection{Ablation Study}

We further conduct ablation studies to understand the contribution of core design choices in both anomaly detection and failure diagnosis.

\begin{table}[htbp]
	\centering
	\setlength{\tabcolsep}{6.4pt}
	\caption{Ablation Study of RFT-FM on Anomaly Detection}
	\label{tab:ablation-anomaly-detection}
	\begin{tabular}{lccc|ccc}
		\toprule
		\multirow{2}{*}{Method} & \multicolumn{3}{c|}{Easy} & \multicolumn{3}{c}{Hard} \\
		\cmidrule(lr){2-4} \cmidrule(lr){5-7}
		& P (\%) & R (\%) & F1 (\%) & P (\%) & R (\%) & F1 (\%) \\
		\midrule
		\textbf{RFT-FM} & \underline{99.60} & \textbf{78.75} & \textbf{87.96} & \underline{99.62} & \textbf{58.71} & \textbf{73.88} \\
		\midrule
		w/o NPC & \textbf{100.00} & 11.87 & 21.23 & \textbf{100.00} & 1.12 & 2.21 \\
		w/o DIS & 98.21 & 17.19 & 29.26 & 96.77 & 6.70 & 12.53 \\
		\bottomrule
	\end{tabular}
\end{table}

Table~\ref{tab:ablation-anomaly-detection} reports the ablation results for anomaly detection. Removing \textit{Normal-Profile Calibration} (NPC) causes a dramatic performance collapse under both easy and hard settings: although precision remains extremely high, recall drops from 78.75\% to 11.87\% on easy faults and from 58.71\% to 1.12\% on hard faults, leading to F1 scores of only 21.23\% and 2.21\%, respectively. This result shows that anomaly detection in RFT is fundamentally relative rather than absolute, and that a calibrated healthy training profile is essential for identifying meaningful deviations. Removing \textit{Deviation-Based Invariant Scoring} (DIS) also severely hurts performance, reducing F1 to 29.26\% on easy faults and 12.53\% on hard faults. This indicates that directly relying on raw statistical features is insufficient, and that structured invariant-based deviation modeling is necessary for robust RFT anomaly detection. Overall, the full RFT-FM detector substantially outperforms both ablated variants, confirming that NPC and DIS are both indispensable components.

\begin{table}[htbp]
	\centering
	\setlength{\tabcolsep}{5pt}
	\caption{Ablation Study of RFT-FM on Failure Diagnosis}
	\label{tab:ablation-anomaly-diagnosis}
	\begin{tabular}{lccc|ccc}
		\toprule
		\multirow{2}{*}{Method} & \multicolumn{3}{c|}{Easy} & \multicolumn{3}{c}{Hard} \\
		\cmidrule(lr){2-4} \cmidrule(lr){5-7}
		& P (\%) & R (\%) & F1 (\%) & P (\%) & R (\%) & F1 (\%) \\
		\midrule
		\textbf{RFT-FM} & \textbf{88.37} & \textbf{85.94} & \textbf{85.51} & \textbf{53.37} & \textbf{39.06} & \textbf{42.16} \\
		\midrule
		w/o TD & 83.85 & 82.50 & 80.88 & 42.86 & 35.49 & 35.86 \\
		w/o FGF & 72.37 & 73.12 & 71.70 & 32.72 & 29.91 & 29.03 \\
		w/o TD \& FGF & 74.98 & 75.63 & 74.34 & 54.17 & 29.02 & 31.76 \\
		\bottomrule
	\end{tabular}
\end{table}

Table~\ref{tab:ablation-anomaly-diagnosis} reports the ablation results for failure diagnosis. Removing \textit{Temporal Dynamics Modeling} (TD) degrades type-level diagnosis from 85.51\% to 80.88\% F1 on easy faults and from 42.16\% to 35.86\% on hard faults, showing that fine-grained fault attribution depends on how anomalies unfold over training rather than only on static summaries. Removing \textit{Fine-Grained Fault Fingerprinting} (FGF) causes an even larger degradation, especially under the hard setting, where F1 drops to 29.03\%. This suggests that subtle fault diagnosis requires richer structured fingerprints than those used by anomaly detection alone. When both TD and FGF are removed, diagnosis performance further deteriorates overall, particularly under the hard setting, confirming that the combination of temporal modeling and fine-grained fingerprinting is critical for robust fault attribution.

Overall, the ablation results support the modular design of RFT-FM. For anomaly detection, reliable scoring depends on both calibrated healthy references and invariant-based deviation modeling. For failure diagnosis, robust attribution depends on both temporal dynamics and fine-grained fault fingerprints. These findings are consistent with the empirical study: RFT failures are not only observable, but also structurally distinguishable, and effective failure management therefore requires explicitly modeling both properties.

\subsection{Hyperparameter Analysis}

\begin{figure}[htbp]
	\centering
	\subfigure[Detection]{
		\begin{minipage}{0.46\linewidth}
			\centering
			\includegraphics[width=\linewidth]{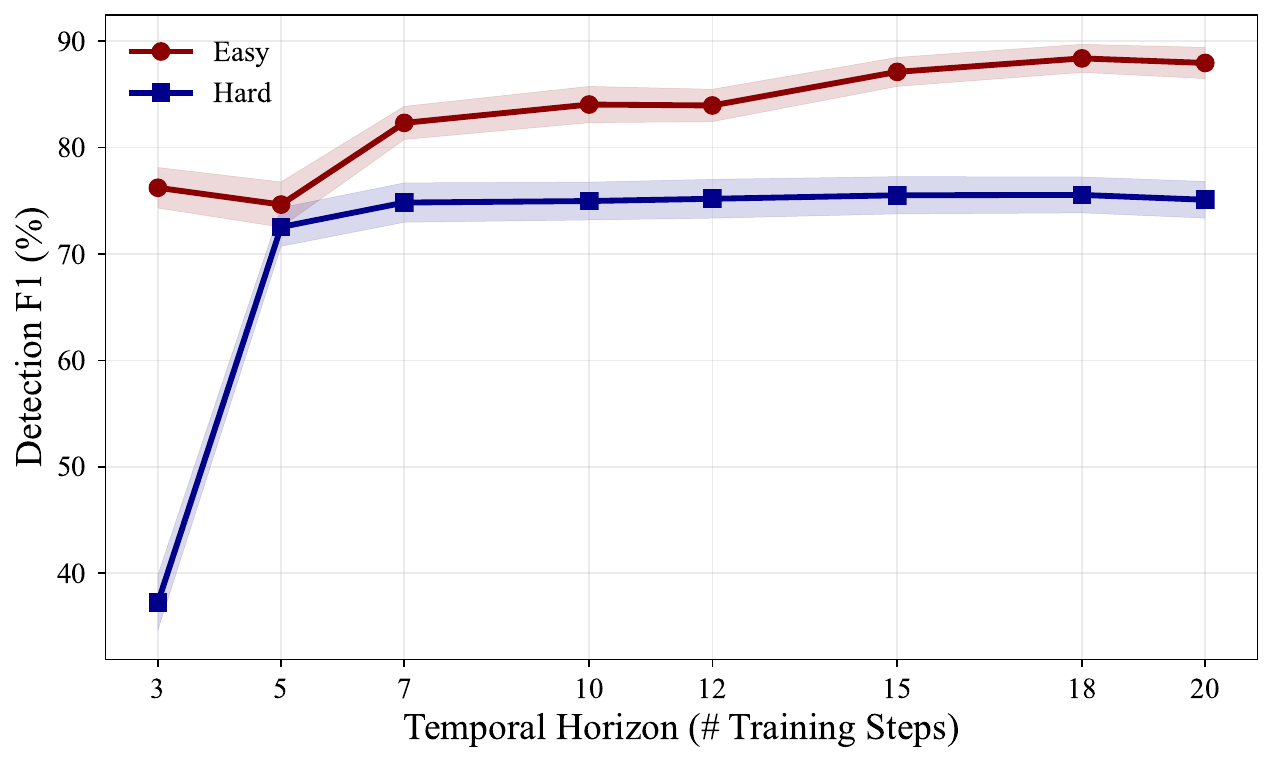}
			\label{fig:detection-sensitivity}
		\end{minipage}
	}
	\subfigure[Diagnosis]{
		\begin{minipage}{0.46\linewidth}
			\centering
			\includegraphics[width=\linewidth]{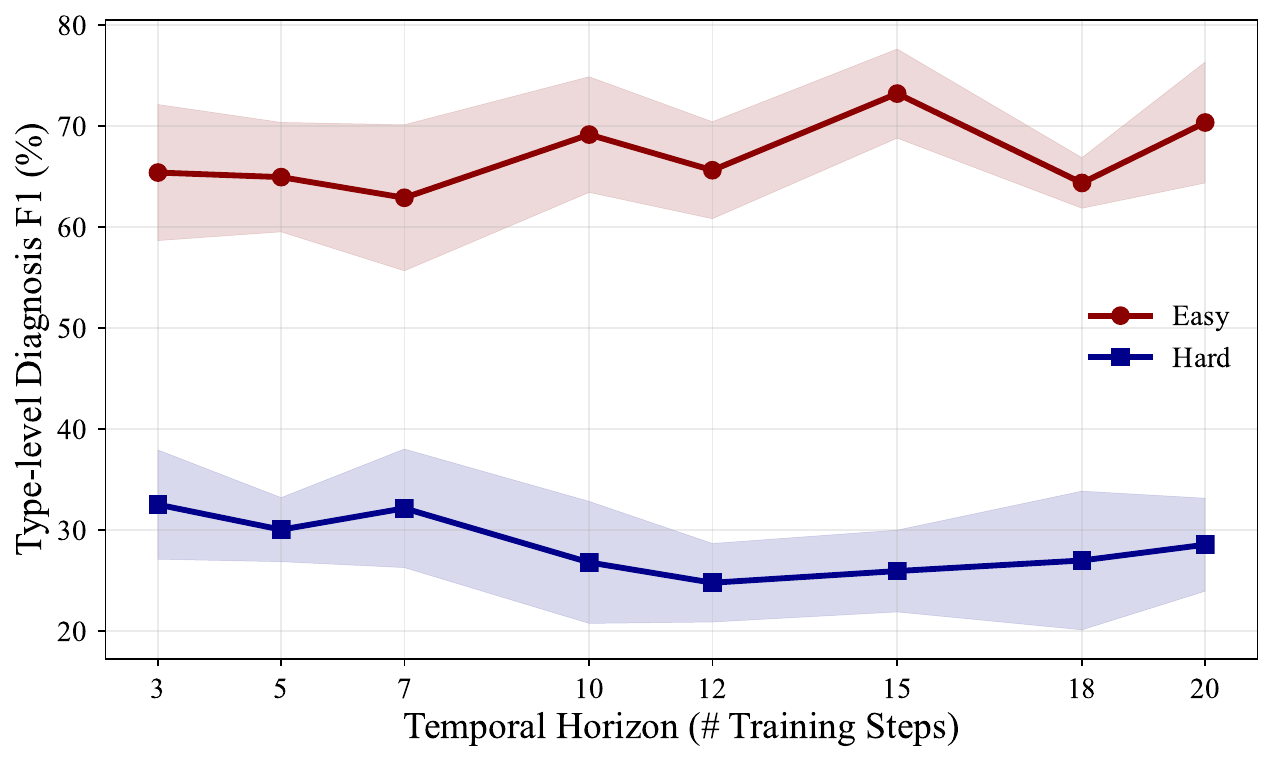}
			\label{fig:diagnosis-sensitivity}
		\end{minipage}
	}
	\caption{Hyperparameter Experiment}
	\label{fig:hyperparameter-sensitivity}
\end{figure}

We further analyze the sensitivity of RFT-FM to one key hyperparameter, namely the temporal horizon used to summarize training dynamics. Figure~\ref{fig:hyperparameter-sensitivity} reports the effect of varying the temporal horizon on both anomaly detection and failure diagnosis.

As shown in Figure~\ref{fig:detection-sensitivity}, anomaly detection performance generally improves as the temporal horizon increases. Under the easy setting, detection F1 rises steadily from shorter horizons and reaches its strongest performance around 18--20 steps, indicating that longer observation windows provide more complete evidence for identifying structured deviations from healthy RFT dynamics. Under the hard setting, the gains are much smaller after the initial increase, suggesting that subtle anomalies expose weaker observable signals and therefore saturate earlier even when more steps are included.

Figure~\ref{fig:diagnosis-sensitivity} shows a similar but more unstable trend for failure diagnosis. Under the easy setting, type-level diagnosis benefits from moderately longer horizons and reaches its best performance around the middle-to-late range, indicating that fine-grained fault attribution requires sufficient temporal context to reveal discriminative fault fingerprints. Under the hard setting, diagnosis remains substantially more difficult across all horizons and exhibits larger fluctuations, showing that subtle failures are not only harder to detect but also much harder to attribute reliably at a fine-grained level.

Overall, the hyperparameter results suggest that temporal horizon plays a meaningful role in both detection and diagnosis. A horizon that is too short may miss critical fault evolution patterns, while a sufficiently long horizon enables RFT-FM to better capture both anomaly severity and fault-specific temporal fingerprints. At the same time, the weaker gains under the hard setting further confirm that subtle RFT failures remain intrinsically challenging even with extended temporal context.

\section{Related Work}

\subsection{Robust Reinforcement Fine-Tuning}

Recent efforts on robust reinforcement fine-tuning mainly improve reliability from two perspectives: system-level robustness and algorithm-level stabilization.

\textbf{System-level robustness.}
One line of work focuses on improving the reliability of large-scale LLM post-training under infrastructure and execution failures. For example, L4~\cite{jiang2025l4} diagnoses large-scale LLM training failures by exploiting cross-job, spatial, and temporal log patterns to localize faulty nodes and failure-indicating logs. RobustRL~\cite{chen2025role} improves the Effective Training Time Ratio (ETTR) of RL post-training under GPU machine failures through a Detect-Restart-Reconnect paradigm. Other system efforts such as TRANSOM~\cite{wu2023transom} and FlashRecovery~\cite{zhang2025flashrecovery} further improve recovery efficiency through automated monitoring, restart orchestration, and low-cost restoration mechanisms. These studies primarily enhance the robustness of the training infrastructure and execution environment, rather than characterizing the internal failure patterns of the fine-tuning process itself.

\textbf{Algorithm-level stabilization.}
Another line of work focuses on mitigating instability within RLHF or RFT by modifying learning objectives, reward modeling, or optimization strategies. Prior studies investigate reward shaping and reward-model redesign to reduce reward hacking and improve training stability~\cite{fu2025reward, miao2025information, duan2026mitigating}. Other work revisits KL regularization and stable optimization in RLHF to better balance alignment quality and training robustness~\cite{liu2025rethinking, heunifying}. These methods improve specific failure modes or stabilization issues in individual subproblems of the RFT pipeline, such as reward exploitation or unstable policy updates.

In contrast to the above directions, our work studies robustness from the perspective of \emph{training-process-level failure management}. Rather than only protecting infrastructure or modifying individual algorithmic components, we explicitly model RFT failures as a structured anomaly space, construct a benchmark for fine-grained RFT failures, and develop a unified framework for anomaly detection, failure diagnosis, and automatic remediation.

\subsection{Software Failure Management}

Software failure management has been widely studied for many years. From a task perspective, it is highly related to our problem setting and can generally be organized into three closely related directions: anomaly detection, failure diagnosis, and automatic remediation.

\textbf{Anomaly detection.}
A large body of work aims to identify abnormal system behaviors from runtime signals such as metrics, logs, and traces. Early studies mainly rely on statistical thresholding and rule-based monitoring~\cite{du2017deeplog, liu2012isolation, song2007conditional}, which are effective for coarse-grained abnormality screening but often struggle with subtle or multivariate failure patterns. More recent approaches adopt machine learning and deep learning models to detect complex anomalies from large-scale operational data~\cite{zhang2024multivariate, zhang2024reducing, zhang2025log, liu2025aaad, duan2025logaction, he2025walk}. With the emergence of large language models, several studies further explore LLM-based anomaly detection~\cite{zhang2025xraglog, xiao2025coorlog, zhang2026runtimeslicer, yang2025ad, liu2025ora}, using stronger semantic reasoning over logs and telemetry. However, due to the inference latency and deployment cost of LLMs, such approaches are still less common in practical anomaly detection pipelines.

\textbf{Failure diagnosis.}
Once anomalies are detected, failure diagnosis seeks to localize root causes and explain why failures occur. Representative directions include failure category classification and root-cause localization. Failure category classification is often formulated as a multi-class classification problem, where anomalies are mapped to predefined failure categories such as resource bottlenecks, configuration errors, or software faults~\cite{zhang2024lm, chen2024automatic, zhang2025scalalog, zhang2025agentfm, sun2023test, gupta2023learning, sun2024art, zhang2024towards, hong2025cslparser, he2025united, zhang2026efficient}. Root-cause localization, in contrast, focuses on identifying the specific service, component, or machine responsible for the failure, often through dependency analysis, causal discovery, or topology-aware reasoning~\cite{zhou2019latent, gan2019seer, liu2020unsupervised, gan2021sage, yu2021microrank, yu2023tracerank, zhang2024trace, zhang2025thinkfl, zhang2025adaptive, zhang2026hypothesize, zhang2026agentic, huang2025uda}. Compared with traditional software failure diagnosis, our setting is closer to failure category classification, but the diagnosis target is fine-grained RFT training failures reflected in training dynamics rather than conventional runtime incidents.

\textbf{Automatic remediation.}
Automatic remediation studies how to recover faulty systems after diagnosis, typically through mitigation solution generation, remediation script generation, and closed-loop recovery execution. Recent work explores LLM-based mitigation planning from incident reports and operational evidence~\cite{ahmed2023recommending, goel2024x, hamadanian2023holistic}, aiming to recommend actionable recovery strategies for on-call engineers. Another line of work focuses on generating executable artifacts such as shell commands, KQL queries, or Ansible playbooks to directly automate repair actions~\cite{jiang2024xpert, shi2023shellgpt, pujar2023automated, sarda2024leveraging, sahoo2024ansible}. More recently, some studies begin to explore end-to-end recovery pipelines that directly connect diagnosis with executable recovery actions~\cite{zhang2026e2e, zhang2025microremed}. In contrast to these works, our study focuses on automatic remediation for reinforcement fine-tuning, where the objective is not to recover running software services, but to mitigate failures arising in the training process itself.

\section{Conclusion}

In this paper, we study the problem of failure management in reinforcement fine-tuning. To enable systematic analysis and evaluation, we construct RFT-FaultBench, the first benchmark for fine-grained RFT failures, covering diverse fault families, fault types, and both easy and hard anomaly settings. Based on this benchmark, we conduct a comprehensive empirical study showing that RFT failures are both observable from training dynamics and distinguishable through structured anomaly patterns. Building on these findings, we propose RFT-FM, an automatic failure management framework for reinforcement fine-tuning. Experimental results show that RFT-FaultBench poses substantial challenges for both generic anomaly-analysis methods and structured diagnosis baselines, while RFT-FM achieves strong performance in detecting, diagnosing, and mitigating RFT failures.

\section*{Acknowledgment}

This work is supported by Key RD Project of Guangdong Province, China (No.2020B010164003).

\balance
\bibliographystyle{IEEEtran}
\bibliography{mylib}

@String{Computing = "Computing" }

@String{Computer = "{IEEE} Computer" }

@String{Chelsea = "Chelsea" }

@article{zhang2025survey,
	title={A Survey of AIOps in the Era of Large Language Models},
	author={Zhang, Lingzhe and Jia, Tong and Jia, Mengxi and Wu, Yifan and Liu, Aiwei and Yang, Yong and Wu, Zhonghai and Hu, Xuming and Yu, Philip and Li, Ying},
	journal={ACM Computing Surveys},
	year={2025},
	publisher={ACM New York, NY}
}

@article{zhang2025adaptive,
	title={Adaptive root cause localization for microservice systems with multi-agent recursion-of-thought},
	author={Zhang, Lingzhe and Jia, Tong and Wang, Kangjin and Hong, Weijie and Duan, Chiming and He, Minghua and Li, Ying},
	journal={arXiv preprint arXiv:2508.20370},
	year={2025}
}

@article{notaro2021survey,
	title={A survey of aiops methods for failure management},
	author={Notaro, Paolo and Cardoso, Jorge and Gerndt, Michael},
	journal={ACM Transactions on Intelligent Systems and Technology (TIST)},
	volume={12},
	number={6},
	pages={1--45},
	year={2021},
	publisher={ACM New York, NY}
}

@article{remil2024aiops,
	title={Aiops solutions for incident management: Technical guidelines and a comprehensive literature review},
	author={Remil, Youcef and Bendimerad, Anes and Mathonat, Romain and Kaytoue, Mehdi},
	journal={arXiv preprint arXiv:2404.01363},
	year={2024}
}

@article{guo2025deepseek,
	title={Deepseek-r1 incentivizes reasoning in llms through reinforcement learning},
	author={Guo, Daya and Yang, Dejian and Zhang, Haowei and Song, Junxiao and Wang, Peiyi and Zhu, Qihao and Xu, Runxin and Zhang, Ruoyu and Ma, Shirong and Bi, Xiao and others},
	journal={Nature},
	volume={645},
	number={8081},
	pages={633--638},
	year={2025},
	publisher={Nature Publishing Group UK London}
}

@article{el2025competitive,
	title={Competitive programming with large reasoning models},
	author={El-Kishky, Ahmed and Wei, Alexander and Saraiva, Andre and Minaiev, Borys and Selsam, Daniel and Dohan, David and Song, Francis and Lightman, Hunter and Clavera, Ignasi and Pachocki, Jakub and others},
	journal={arXiv preprint arXiv:2502.06807},
	year={2025}
}

@inproceedings{jiang2025l4,
	title={L4: Diagnosing large-scale llm training failures via automated log analysis},
	author={Jiang, Zhihan and Huang, Junjie and Yu, Guangba and Chen, Zhuangbin and Li, Yichen and Zhong, Renyi and Feng, Cong and Yang, Yongqiang and Yang, Zengyin and Lyu, Michael},
	booktitle={Proceedings of the 33rd ACM International Conference on the Foundations of Software Engineering},
	pages={51--63},
	year={2025}
}

@article{chen2025role,
	title={Role-Based Fault Tolerance System for LLM RL Post-Training},
	author={Chen, Zhenqian and Zhong, Baoquan and Li, Xiang and Dai, Qing and Zhao, Xinkui and Ye, Miao and Cheng, Ren and Zhang, Lufei and Yin, Jianwei},
	journal={arXiv preprint arXiv:2512.22492},
	year={2025}
}

@article{wu2023transom,
	title={Transom: An efficient fault-tolerant system for training llms},
	author={Wu, Baodong and Xia, Lei and Li, Qingping and Li, Kangyu and Chen, Xu and Guo, Yongqiang and Xiang, Tieyao and Chen, Yuheng and Li, Shigang},
	journal={arXiv preprint arXiv:2310.10046},
	year={2023}
}

@article{zhang2025flashrecovery,
	title={Flashrecovery: Fast and low-cost recovery from failures for large-scale training of llms},
	author={Zhang, Haijun and Wang, Jinxiang and Yu, Zhenhua and Zhang, Yanyong and Ji, Xuejie and Mao, Kaining and Zhang, Jun and Zhang, Yaqing and Wu, Ting and Jie, Fei and others},
	journal={arXiv preprint arXiv:2509.03047},
	year={2025}
}

@article{fu2025reward,
	title={Reward shaping to mitigate reward hacking in rlhf},
	author={Fu, Jiayi and Zhao, Xuandong and Yao, Chengyuan and Wang, Heng and Han, Qi and Xiao, Yanghua},
	journal={arXiv preprint arXiv:2502.18770},
	year={2025}
}

@article{duan2026mitigating,
	title={Mitigating Reward Hacking in RLHF via Bayesian Non-negative Reward Modeling},
	author={Duan, Zhibin and Rong, Guowei and Li, Zhuo and Chen, Bo and Zhou, Mingyuan and Guo, Dandan},
	journal={arXiv preprint arXiv:2602.10623},
	year={2026}
}

@article{liu2025rethinking,
	title={Rethinking kl regularization in rlhf: From value estimation to gradient optimization},
	author={Liu, Kezhao and Liu, Jason Klein and Chen, Mingtao and Liu, Yiming},
	journal={arXiv preprint arXiv:2510.01555},
	year={2025}
}

@article{zhang2025thinkfl,
	title={Thinkfl: Self-refining failure localization for microservice systems via reinforcement fine-tuning},
	author={Zhang, Lingzhe and Zhai, Yunpeng and Jia, Tong and Duan, Chiming and Yu, Siyu and Gao, Jinyang and Ding, Bolin and Wu, Zhonghai and Li, Ying},
	journal={ACM Transactions on Software Engineering and Methodology},
	year={2025},
	publisher={ACM New York, NY}
}

@inproceedings{heunifying,
	title={Unifying Stable Optimization and Reference Regularization in RLHF},
	author={He, Li and Qu, Qiang and Zhao, He and Wan, Stephen and Wang, Dadong and Yao, Lina and Liu, Tongliang},
	booktitle={The Fourteenth International Conference on Learning Representations}
}

@inproceedings{liu2025ora,
	title={ORA: Job Runtime Prediction for High-Performance Computing Platforms Using the Online Retrieval-Augmented Language Model},
	author={Liu, Hongyi and Ma, Yinping and Huang, Xiaosong and Zhang, Lingzhe and Jia, Tong and Li, Ying},
	booktitle={Proceedings of the 39th ACM International Conference on Supercomputing},
	pages={884--894},
	year={2025}
}

@article{christiano2017deep,
	title={Deep reinforcement learning from human preferences},
	author={Christiano, Paul F and Leike, Jan and Brown, Tom and Martic, Miljan and Legg, Shane and Amodei, Dario},
	journal={Advances in neural information processing systems},
	volume={30},
	year={2017}
}

@article{ziegler2019fine,
	title={Fine-tuning language models from human preferences},
	author={Ziegler, Daniel M and Stiennon, Nisan and Wu, Jeffrey and Brown, Tom B and Radford, Alec and Amodei, Dario and Christiano, Paul and Irving, Geoffrey},
	journal={arXiv preprint arXiv:1909.08593},
	year={2019}
}

@article{rafailov2023direct,
	title={Direct preference optimization: Your language model is secretly a reward model},
	author={Rafailov, Rafael and Sharma, Archit and Mitchell, Eric and Manning, Christopher D and Ermon, Stefano and Finn, Chelsea},
	journal={Advances in Neural Information Processing Systems},
	volume={36},
	pages={53728--53741},
	year={2023}
}

@article{schulman2017proximal,
	title={Proximal policy optimization algorithms},
	author={Schulman, John and Wolski, Filip and Dhariwal, Prafulla and Radford, Alec and Klimov, Oleg},
	journal={arXiv preprint arXiv:1707.06347},
	year={2017}
}

@article{shao2024deepseekmath,
	title={Deepseekmath: Pushing the limits of mathematical reasoning in open language models},
	author={Shao, Zhihong and Wang, Peiyi and Zhu, Qihao and Xu, Runxin and Song, Junxiao and Bi, Xiao and Zhang, Haowei and Zhang, Mingchuan and Li, YK and Wu, Y and others},
	journal={arXiv preprint arXiv:2402.03300},
	year={2024}
}

@article{hu2024openrlhf,
	title={Openrlhf: An easy-to-use, scalable and high-performance rlhf framework},
	author={Hu, Jian and Wu, Xibin and Zhu, Zilin and Wang, Weixun and Zhang, Dehao and Cao, Yu and others},
	journal={arXiv preprint arXiv:2405.11143},
	year={2024}
}

@article{zhang2024towards,
	title={Towards close-to-zero runtime collection overhead: Raft-based anomaly diagnosis on system faults for distributed storage system},
	author={Zhang, Lingzhe and Jia, Tong and Jia, Mengxi and Liu, Hongyi and Yang, Yong and Wu, Zhonghai and Li, Ying},
	journal={IEEE Transactions on Services Computing},
	volume={18},
	number={2},
	pages={1054--1067},
	year={2024},
	publisher={IEEE}
}

@inproceedings{wenlanguage,
	title={Language Models Learn to Mislead Humans via RLHF},
	author={Wen, Jiaxin and Zhong, Ruiqi and Khan, Akbir and Perez, Ethan and Steinhardt, Jacob and Huang, Minlie and Bowman, Samuel R and He, He and Feng, Shi},
	booktitle={The Thirteenth International Conference on Learning Representations}
}

@inproceedings{chen2024odin,
	title={ODIN: disentangled reward mitigates hacking in RLHF},
	author={Chen, Lichang and Zhu, Chen and Chen, Jiuhai and Soselia, Davit and Zhou, Tianyi and Goldstein, Tom and Huang, Heng and Shoeybi, Mohammad and Catanzaro, Bryan},
	booktitle={Proceedings of the 41st International Conference on Machine Learning},
	pages={7935--7952},
	year={2024}
}

@article{miao2025information,
	title={Information-Theoretic Reward Modeling for Stable RLHF: Detecting and Mitigating Reward Hacking},
	author={Miao, Yuchun and Ding, Liang and Zhang, Sen and Bao, Rong and Zhang, Lefei and Tao, Dacheng},
	journal={arXiv preprint arXiv:2510.13694},
	year={2025}
}

@inproceedings{ji2025language,
	title={Language models resist alignment: Evidence from data compression},
	author={Ji, Jiaming and Wang, Kaile and Qiu, Tianyi Alex and Chen, Boyuan and Zhou, Jiayi and Li, Changye and Lou, Hantao and Dai, Josef and Liu, Yunhuai and Yang, Yaodong},
	booktitle={Proceedings of the 63rd Annual Meeting of the Association for Computational Linguistics (Volume 1: Long Papers)},
	pages={23411--23432},
	year={2025}
}

@inproceedings{sheshadrisome,
	title={Why Do Some Language Models Fake Alignment While Others Don't?},
	author={Sheshadri, Abhay and Hughes, John and Michael, Julian and Mallen, Alex Troy and Jose, Arun and Roger, Fabien},
	booktitle={The Thirty-ninth Annual Conference on Neural Information Processing Systems}
}

@article{xue2026supervised,
	title={Why Supervised Fine-Tuning Fails to Learn: A Systematic Study of Incomplete Learning in Large Language Models},
	author={Xue, Chao and Wang, Yao and Liu, Mengqiao and Liang, Di and Han, Xingsheng and Liu, Peiyang and Wu, Xianjie and Lu, Chenyao and Jiang, Lei and Lu, Yu and others},
	journal={arXiv preprint arXiv:2604.10079},
	year={2026}
}

@article{ouyang2022training,
	title={Training language models to follow instructions with human feedback},
	author={Ouyang, Long and Wu, Jeffrey and Jiang, Xu and Almeida, Diogo and Wainwright, Carroll and Mishkin, Pamela and Zhang, Chong and Agarwal, Sandhini and Slama, Katarina and Ray, Alex and others},
	journal={Advances in neural information processing systems},
	volume={35},
	pages={27730--27744},
	year={2022}
}

@article{tuli2022tranad,
	title={TranAD: deep transformer networks for anomaly detection in multivariate time series data},
	author={Tuli, Shreshth and Casale, Giuliano and Jennings, Nicholas R},
	journal={Proceedings of the VLDB Endowment},
	volume={15},
	number={6},
	pages={1201--1214},
	year={2022},
	publisher={VLDB Endowment}
}

@inproceedings{su2019robust,
	title={Robust anomaly detection for multivariate time series through stochastic recurrent neural network},
	author={Su, Ya and Zhao, Youjian and Niu, Chenhao and Liu, Rong and Sun, Wei and Pei, Dan},
	booktitle={Proceedings of the 25th ACM SIGKDD international conference on knowledge discovery \& data mining},
	pages={2828--2837},
	year={2019}
}

@inproceedings{xuanomaly,
	title={Anomaly Transformer: Time Series Anomaly Detection with Association Discrepancy},
	author={Xu, Jiehui and Wu, Haixu and Wang, Jianmin and Long, Mingsheng},
	booktitle={International Conference on Learning Representations}
}

@inproceedings{lin2024root,
	title={Root Cause Analysis in Microservice Using Neural Granger Causal Discovery},
	author={Lin, Cheng-Ming and Chang, Ching and Wang, Wei-Yao and Wang, Kuang-Da and Peng, Wen-Chih},
	booktitle={Proceedings of the AAAI Conference on Artificial Intelligence},
	volume={38},
	number={1},
	pages={206--213},
	year={2024}
}

@article{xin2023causalrca,
	title={Causalrca: Causal inference based precise fine-grained root cause localization for microservice applications},
	author={Xin, Ruyue and Chen, Peng and Zhao, Zhiming},
	journal={Journal of Systems and Software},
	volume={203},
	pages={111724},
	year={2023},
	publisher={Elsevier}
}

@article{jiang2025circa,
	title={CIRCA: A Framework for Collaborative Identification of Root Cause Analysis in IoT Microservices},
	author={Jiang, Xingguo and Luo, Hong and Sun, Yan and Das, Sajal K},
	journal={IEEE Transactions on Services Computing},
	year={2025},
	publisher={IEEE}
}

@article{zhang2026e2e,
	title={E2E-REME: Towards End-to-End Microservices Auto-Remediation via Experience-Simulation Reinforcement Fine-Tuning},
	author={Zhang, Lingzhe and Zhai, Yunpeng and Jia, Tong and He, Minghua and Duan, Chiming and Liu, Zhaoyang and Ding, Bolin and Li, Ying},
	journal={arXiv preprint arXiv:2604.11094},
	year={2026}
}

@article{pan2025d,
	title={d-TreeRPO: Towards More Reliable Policy Optimization for Diffusion Language Models},
	author={Pan, Leyi and Tao, Shuchang and Zhai, Yunpeng and Fu, Zheyu and Fang, Liancheng and He, Minghua and Zhang, Lingzhe and Liu, Zhaoyang and Ding, Bolin and Liu, Aiwei and others},
	journal={arXiv preprint arXiv:2512.09675},
	year={2025}
}

@article{zhang2025microremed,
	title={MicroRemed: Benchmarking LLMs in Microservices Remediation},
	author={Zhang, Lingzhe and Zhai, Yunpeng and Jia, Tong and Duan, Chiming and He, Minghua and Pan, Leyi and Liu, Zhaoyang and Ding, Bolin and Li, Ying},
	journal={arXiv preprint arXiv:2511.01166},
	year={2025}
}

@article{pan2025omni,
	title={Omni-SafetyBench: A benchmark for safety evaluation of audio-visual large language models},
	author={Pan, Leyi and Fu, Zheyu and Zhai, Yunpeng and Tao, Shuchang and Guan, Sheng and Huang, Shiyu and Zhang, Lingzhe and Liu, Zhaoyang and Ding, Bolin and Henry, Felix and others},
	journal={arXiv preprint arXiv:2508.07173},
	year={2025}
}

@article{liu2012isolation,
	title={Isolation-based anomaly detection},
	author={Liu, Fei Tony and Ting, Kai Ming and Zhou, Zhi-Hua},
	journal={ACM Transactions on Knowledge Discovery from Data (TKDD)},
	volume={6},
	number={1},
	pages={1--39},
	year={2012},
	publisher={ACM New York, NY, USA}
}

@article{song2007conditional,
	title={Conditional anomaly detection},
	author={Song, Xiuyao and Wu, Mingxi and Jermaine, Christopher and Ranka, Sanjay},
	journal={IEEE Transactions on knowledge and Data Engineering},
	volume={19},
	number={5},
	pages={631--645},
	year={2007},
	publisher={IEEE}
}

@inproceedings{zhang2024multivariate,
	title={Multivariate log-based anomaly detection for distributed database},
	author={Zhang, Lingzhe and Jia, Tong and Jia, Mengxi and Li, Ying and Yang, Yong and Wu, Zhonghai},
	booktitle={Proceedings of the 30th ACM SIGKDD Conference on Knowledge Discovery and Data Mining},
	pages={4256--4267},
	year={2024}
}

@inproceedings{zhang2024reducing,
	title={Reducing events to augment log-based anomaly detection models: An empirical study},
	author={Zhang, Lingzhe and Jia, Tong and Wang, Kangjin and Jia, Mengxi and Yang, Yong and Li, Ying},
	booktitle={Proceedings of the 18th ACM/IEEE International Symposium on Empirical Software Engineering and Measurement},
	pages={538--548},
	year={2024}
}

@inproceedings{he2025walk,
	title={Walk the Talk: Is Your Log-based Software Reliability Maintenance System Really Reliable?},
	author={He, Minghua and Jia, Tong and Duan, Chiming and Xiao, Pei and Zhang, Lingzhe and Wang, Kangjin and Wu, Yifan and Li, Ying and Huang, Gang},
	booktitle={2025 40th IEEE/ACM International Conference on Automated Software Engineering (ASE)},
	pages={3784--3788},
	year={2025},
	organization={IEEE}
}

@article{zhang2025log,
	title={E-log: Fine-grained elastic log-based anomaly detection and diagnosis for databases},
	author={Zhang, Lingzhe and Jia, Tong and Tan, Xinyu and Huang, Xiangdong and Jia, Mengxi and Liu, Hongyi and Wu, Zhonghai and Li, Ying},
	journal={IEEE Transactions on Services Computing},
	year={2025},
	publisher={IEEE}
}

@inproceedings{liu2025aaad,
	title={AAAD: Asynchronous Inter-Variable Relationship-Aware Anomaly Detection for Multivariate Time Series},
	author={Liu, Hongyi and Huang, Xiaosong and Jia, Mengxi and Zhang, Lingzhe and Jia, Tong and Wu, Zhonghai and Li, Ying},
	booktitle={2025 IEEE International Conference on Multimedia and Expo (ICME)},
	pages={1--6},
	year={2025},
	organization={IEEE}
}

@inproceedings{zhang2025xraglog,
	title={XRAGLog: A resource-efficient and context-aware log-based anomaly detection method using retrieval-augmented generation},
	author={Zhang, Lingzhe and Jia, Tong and Jia, Mengxi and Wu, Yifan and Liu, Hongyi and Li, Ying},
	booktitle={AAAI 2025 Workshop on Preventing and Detecting LLM Misinformation (PDLM)},
	year={2025}
}

@inproceedings{duan2025logaction,
	title={LogAction: Consistent Cross-system Anomaly Detection through Logs via Active Domain Adaptation},
	author={Duan, Chiming and He, Minghua and Xiao, Pei and Jia, Tong and Zhang, Xin and Zhong, Zhewei and Luo, Xiang and Niu, Yan and Zhang, Lingzhe and Yu, Siyu and others},
	booktitle={2025 40th IEEE/ACM International Conference on Automated Software Engineering (ASE)},
	pages={700--712},
	year={2025},
	organization={IEEE}
}

@inproceedings{xiao2025coorlog,
	title={CoorLog: Efficient-Generalizable Log Anomaly Detection via Adaptive Coordinator in Software Evolution},
	author={Xiao, Pei and Duan, Chiming and He, Minghua and Jia, Tong and Wu, Yifan and Xu, Jing and Gao, Gege and Zhang, Lingzhe and Hong, Weijie and Li, Ying and others},
	booktitle={2025 40th IEEE/ACM International Conference on Automated Software Engineering (ASE)},
	pages={1119--1131},
	year={2025},
	organization={IEEE}
}

@article{zhang2026runtimeslicer,
	title={RuntimeSlicer: Towards Generalizable Unified Runtime State Representation for Failure Management},
	author={Zhang, Lingzhe and Jia, Tong and Hong, Weijie and Wang, Mingyu and Duan, Chiming and He, Minghua and Wang, Rongqian and Peng, Xi and Wang, Meiling and Zhang, Gong and others},
	journal={arXiv preprint arXiv:2603.21495},
	year={2026}
}

@inproceedings{yang2025ad,
	title={Ad-llm: Benchmarking large language models for anomaly detection},
	author={Yang, Tiankai and Nian, Yi and Li, Li and Xu, Ruiyao and Li, Yuangang and Li, Jiaqi and Xiao, Zhuo and Hu, Xiyang and Rossi, Ryan A and Ding, Kaize and others},
	booktitle={Findings of the Association for Computational Linguistics: ACL 2025},
	pages={1524--1547},
	year={2025}
}

@inproceedings{zhang2024lm,
	title={LM-PACE: Confidence estimation by large language models for effective root causing of cloud incidents},
	author={Zhang, Dylan and Zhang, Xuchao and Bansal, Chetan and Las-Casas, Pedro and Fonseca, Rodrigo and Rajmohan, Saravan},
	booktitle={Companion Proceedings of the 32nd ACM International Conference on the Foundations of Software Engineering},
	pages={388--398},
	year={2024}
}

@inproceedings{chen2024automatic,
	title={Automatic root cause analysis via large language models for cloud incidents},
	author={Chen, Yinfang and Xie, Huaibing and Ma, Minghua and Kang, Yu and Gao, Xin and Shi, Liu and Cao, Yunjie and Gao, Xuedong and Fan, Hao and Wen, Ming and others},
	booktitle={Proceedings of the Nineteenth European Conference on Computer Systems},
	pages={674--688},
	year={2024}
}

@article{zhang2025survey2,
	title={A survey on parallel text generation: From parallel decoding to diffusion language models},
	author={Zhang, Lingzhe and Fang, Liancheng and Duan, Chiming and He, Minghua and Pan, Leyi and Xiao, Pei and Huang, Shiyu and Zhai, Yunpeng and Hu, Xuming and Yu, Philip S and others},
	journal={arXiv preprint arXiv:2508.08712},
	year={2025}
}

@inproceedings{zhang2025scalalog,
	title={ScalaLog: Scalable Log-Based Failure Diagnosis Using LLM},
	author={Zhang, Lingzhe and Jia, Tong and Jia, Mengxi and Wu, Yifan and Liu, Hongyi and Li, Ying},
	booktitle={ICASSP 2025-2025 IEEE International Conference on Acoustics, Speech and Signal Processing (ICASSP)},
	pages={1--5},
	year={2025},
	organization={IEEE}
}

@article{zhang2025agentfm,
	title={AgentFM: Role-Aware Failure Management for Distributed Databases with LLM-Driven Multi-Agents},
	author={Zhang, Lingzhe and Zhai, Yunpeng and Jia, Tong and Huang, Xiaosong and Duan, Chiming and Li, Ying},
	journal={arXiv preprint arXiv:2504.06614},
	year={2025}
}

@inproceedings{sun2023test,
	title={TEST: Text Prototype Aligned Embedding to Activate LLM's Ability for Time Series},
	author={Sun, Chenxi and Li, Hongyan and Li, Yaliang and Hong, Shenda},
	booktitle={The Twelfth International Conference on Learning Representations},
	year={2024}
}

@inproceedings{gupta2023learning,
	title={Learning Representations on Logs for AIOps},
	author={Gupta, Pranjal and Kumar, Harshit and Kar, Debanjana and Bhukar, Karan and Aggarwal, Pooja and Mohapatra, Prateeti},
	booktitle={2023 IEEE 16th International Conference on Cloud Computing (CLOUD)},
	pages={155--166},
	year={2023},
	organization={IEEE}
}

@inproceedings{sun2024art,
	title={ART: A Unified Unsupervised Framework for Incident Management in Microservice Systems},
	author={Sun, Yongqian and Shi, Binpeng and Mao, Mingyu and Ma, Minghua and Xia, Sibo and Zhang, Shenglin and Pei, Dan},
	booktitle={Proceedings of the 39th IEEE/ACM International Conference on Automated Software Engineering},
	pages={1183--1194},
	year={2024}
}

@inproceedings{zhou2019latent,
	title={Latent error prediction and fault localization for microservice applications by learning from system trace logs},
	author={Zhou, Xiang and Peng, Xin and Xie, Tao and Sun, Jun and Ji, Chao and Liu, Dewei and Xiang, Qilin and He, Chuan},
	booktitle={Proceedings of the 2019 27th ACM joint meeting on European software engineering conference and symposium on the foundations of software engineering},
	pages={683--694},
	year={2019}
}

@inproceedings{gan2019seer,
	title={Seer: Leveraging big data to navigate the complexity of performance debugging in cloud microservices},
	author={Gan, Yu and Zhang, Yanqi and Hu, Kelvin and Cheng, Dailun and He, Yuan and Pancholi, Meghna and Delimitrou, Christina},
	booktitle={Proceedings of the twenty-fourth international conference on architectural support for programming languages and operating systems},
	pages={19--33},
	year={2019}
}

@inproceedings{liu2020unsupervised,
	title={Unsupervised detection of microservice trace anomalies through service-level deep bayesian networks},
	author={Liu, Ping and Xu, Haowen and Ouyang, Qianyu and Jiao, Rui and Chen, Zhekang and Zhang, Shenglin and Yang, Jiahai and Mo, Linlin and Zeng, Jice and Xue, Wenman and others},
	booktitle={2020 IEEE 31st International Symposium on Software Reliability Engineering (ISSRE)},
	pages={48--58},
	year={2020},
	organization={IEEE}
}

@inproceedings{gan2021sage,
	title={Sage: practical and scalable ML-driven performance debugging in microservices},
	author={Gan, Yu and Liang, Mingyu and Dev, Sundar and Lo, David and Delimitrou, Christina},
	booktitle={Proceedings of the 26th ACM International Conference on Architectural Support for Programming Languages and Operating Systems},
	pages={135--151},
	year={2021}
}

@inproceedings{yu2021microrank,
	title={Microrank: End-to-end latency issue localization with extended spectrum analysis in microservice environments},
	author={Yu, Guangba and Chen, Pengfei and Chen, Hongyang and Guan, Zijie and Huang, Zicheng and Jing, Linxiao and Weng, Tianjun and Sun, Xinmeng and Li, Xiaoyun},
	booktitle={Proceedings of the Web Conference 2021},
	pages={3087--3098},
	year={2021}
}

@article{yu2023tracerank,
	title={TraceRank: Abnormal service localization with dis-aggregated end-to-end tracing data in cloud native systems},
	author={Yu, Guangba and Huang, Zicheng and Chen, Pengfei},
	journal={Journal of Software: Evolution and Process},
	volume={35},
	number={10},
	pages={e2413},
	year={2023},
	publisher={Wiley Online Library}
}

@inproceedings{zhang2024trace,
	title={Trace-based Multi-Dimensional Root Cause Localization of Performance Issues in Microservice Systems},
	author={Zhang, Chenxi and Dong, Zhen and Peng, Xin and Zhang, Bicheng and Chen, Miao},
	booktitle={Proceedings of the IEEE/ACM 46th International Conference on Software Engineering},
	pages={1--12},
	year={2024}
}

@inproceedings{ahmed2023recommending,
	title={Recommending root-cause and mitigation steps for cloud incidents using large language models},
	author={Ahmed, Toufique and Ghosh, Supriyo and Bansal, Chetan and Zimmermann, Thomas and Zhang, Xuchao and Rajmohan, Saravan},
	booktitle={2023 IEEE/ACM 45th International Conference on Software Engineering (ICSE)},
	pages={1737--1749},
	year={2023},
	organization={IEEE}
}

@inproceedings{goel2024x,
	title={X-lifecycle learning for cloud incident management using llms},
	author={Goel, Drishti and Husain, Fiza and Singh, Aditya and Ghosh, Supriyo and Parayil, Anjaly and Bansal, Chetan and Zhang, Xuchao and Rajmohan, Saravan},
	booktitle={Companion Proceedings of the 32nd ACM International Conference on the Foundations of Software Engineering},
	pages={417--428},
	year={2024}
}

@inproceedings{hamadanian2023holistic,
	title={A Holistic View of AI-driven Network Incident Management},
	author={Hamadanian, Pouya and Arzani, Behnaz and Fouladi, Sadjad and Kakarla, Siva Kesava Reddy and Fonseca, Rodrigo and Billor, Denizcan and Cheema, Ahmad and Nkposong, Edet and Chandra, Ranveer},
	booktitle={Proceedings of the 22nd ACM Workshop on Hot Topics in Networks},
	pages={180--188},
	year={2023}
}

@inproceedings{jiang2024xpert,
	title={Xpert: Empowering incident management with query recommendations via large language models},
	author={Jiang, Yuxuan and Zhang, Chaoyun and He, Shilin and Yang, Zhihao and Ma, Minghua and Qin, Si and Kang, Yu and Dang, Yingnong and Rajmohan, Saravan and Lin, Qingwei and others},
	booktitle={Proceedings of the IEEE/ACM 46th International Conference on Software Engineering},
	pages={1--13},
	year={2024}
}

@inproceedings{shi2023shellgpt,
	title={ShellGPT: Generative Pre-trained Transformer Model for Shell Language Understanding},
	author={Shi, Jie and Jiang, Sihang and Xu, Bo and Liang, Jiaqing and Xiao, Yanghua and Wang, Wei},
	booktitle={2023 IEEE 34th International Symposium on Software Reliability Engineering (ISSRE)},
	pages={671--682},
	year={2023},
	organization={IEEE}
}

@inproceedings{pujar2023automated,
	title={Automated code generation for information technology tasks in yaml through large language models},
	author={Pujar, Saurabh and Buratti, Luca and Guo, Xiaojie and Dupuis, Nicolas and Lewis, Burn and Suneja, Sahil and Sood, Atin and Nalawade, Ganesh and Jones, Matt and Morari, Alessandro and others},
	booktitle={2023 60th ACM/IEEE Design Automation Conference (DAC)},
	pages={1--4},
	year={2023},
	organization={IEEE}
}

@inproceedings{sarda2024leveraging,
	title={Leveraging large language models for the auto-remediation of microservice applications: An experimental study},
	author={Sarda, Komal and Namrud, Zakeya and Litoiu, Marin and Shwartz, Larisa and Watts, Ian},
	booktitle={Companion Proceedings of the 32nd ACM International Conference on the Foundations of Software Engineering},
	pages={358--369},
	year={2024}
}

@inproceedings{sahoo2024ansible,
	title={Ansible lightspeed: A code generation service for it automation},
	author={Sahoo, Priyam and Pujar, Saurabh and Nalawade, Ganesh and Genhardt, Richard and Mandel, Louis and Buratti, Luca},
	booktitle={Proceedings of the 39th IEEE/ACM International Conference on Automated Software Engineering},
	pages={2148--2158},
	year={2024}
}

@inproceedings{du2017deeplog,
	title={Deeplog: Anomaly detection and diagnosis from system logs through deep learning},
	author={Du, Min and Li, Feifei and Zheng, Guineng and Srikumar, Vivek},
	booktitle={Proceedings of the 2017 ACM SIGSAC conference on computer and communications security},
	pages={1285--1298},
	year={2017}
}

@inproceedings{hong2025cslparser,
	title={CSLParser: A Collaborative Framework Using Small and Large Language Models for Log Parsing},
	author={Hong, Weijie and Wu, Yifan and Zhang, Lingzhe and Duan, Chiming and Xiao, Pei and He, Minghua and Yang, Xixuan and Li, Ying},
	booktitle={2025 IEEE 36th International Symposium on Software Reliability Engineering (ISSRE)},
	pages={61--72},
	year={2025},
	organization={IEEE}
}

@inproceedings{he2025united,
	title={United We Stand: Towards End-to-End Log-based Fault Diagnosis via Interactive Multi-Task Learning},
	author={He, Minghua and Duan, Chiming and Xiao, Pei and Jia, Tong and Yu, Siyu and Zhang, Lingzhe and Hong, Weijie and Han, Jin and Wu, Yifan and Li, Ying and others},
	booktitle={2025 40th IEEE/ACM International Conference on Automated Software Engineering (ASE)},
	pages={661--673},
	year={2025},
	organization={IEEE}
}

@article{zhang2026hypothesize,
	title={Hypothesize-Then-Verify: Speculative Root Cause Analysis for Microservices with Pathwise Parallelism},
	author={Zhang, Lingzhe and Jia, Tong and Zhai, Yunpeng and Pan, Leyi and Duan, Chiming and He, Minghua and Xiao, Pei and Li, Ying},
	journal={arXiv preprint arXiv:2601.02736},
	year={2026}
}

@article{zhang2026agentic,
	title={Agentic Memory Enhanced Recursive Reasoning for Root Cause Localization in Microservices},
	author={Zhang, Lingzhe and Jia, Tong and Zhai, Yunpeng and Pan, Leyi and Duan, Chiming and He, Minghua and Jia, Mengxi and Li, Ying},
	journal={arXiv preprint arXiv:2601.02732},
	year={2026}
}

@article{huang2025uda,
	title={UDA-RCL: Unsupervised Domain Adaptation for Microservice Root Cause Localization Utilizing Multimodal Data},
	author={Huang, Xiaosong and Liu, Hongyi and Wu, Yifan and Zhang, Lingzhe and Jia, Tong and Li, Ying and Wu, Zhonghai},
	journal={IEEE Transactions on Services Computing},
	year={2025},
	publisher={IEEE}
}

@article{zhang2026efficient,
	title={Efficient Failure Management for Multi-Agent Systems with Reasoning Trace Representation},
	author={Zhang, Lingzhe and Jia, Tong and Wang, Mingyu and Hong, Weijie and Duan, Chiming and He, Minghua and Wang, Rongqian and Peng, Xi and Wang, Meiling and Zhang, Gong and others},
	journal={arXiv preprint arXiv:2603.21522},
	year={2026}
}

\end{document}